\documentclass[11pt,a4paper]{article}
\usepackage{jcappub}

\usepackage{epsfig}
\usepackage{amsmath}
\usepackage{amssymb}
\usepackage{enumerate}
\usepackage{multirow}

\title{
Two radiative inverse seesaw models, dark matter, and baryogenesis.
}

\author[a]{Iason Baldes,}
\author[a]{Nicole F. Bell,}
\author[a,b]{Kalliopi Petraki,}
\author[a]{and \\ Raymond R. Volkas.}
\affiliation[a]{ARC Centre of Excellence for Particle Physics at the Terascale, \\
School of Physics, The University of Melbourne, Victoria 3010, Australia}
\affiliation[b]{Nikhef, Science Park 105, 1098 XG Amsterdam, The Netherlands}

\emailAdd{i.baldes@student.unimelb.edu.au}
\emailAdd{n.bell@unimelb.edu.au}
\emailAdd{kpetraki@nikhef.nl}
\emailAdd{raymondv@unimelb.edu.au}

\abstract{The inverse seesaw mechanism allows the neutrino masses to be generated by new physics at an experimentally accessible scale, even with $\mathcal{O}(1)$ Yukawa couplings. In the inverse seesaw scenario, the smallness of neutrino masses is linked to the smallness of a lepton number violating parameter. This parameter may arise radiatively. In this paper, we study the cosmological implications of two contrasting radiative inverse seesaw models, one due to Ma and the other to Law and McDonald. The former features spontaneous, the latter explicit lepton number violation. First, we examine the effect of the lepton-number violating interactions introduced in these models on the baryon asymmetry of the universe. We investigate under what conditions a pre-existing baryon asymmetry does not get washed out. While both models allow a baryon asymmetry to survive only once the temperature has dropped below the mass of their heaviest fields, the Ma model can create the baryon asymmetry through resonant leptogenesis. Then we investigate the viability of the dark matter candidates arising within these models, and explore the prospects for direct detection. We find that the Law/McDonald model allows a simple dark matter scenario similar to the Higgs portal, while in the Ma model the simplest cold dark matter scenario would tend to overclose the universe.
}
\date{today}
\begin{document}
\maketitle

\section{Introduction}

The experimental evidence for physics beyond the standard model (SM) includes the baryon asymmetry of the universe (BAU), the existence of dark matter (DM), and the non-zero neutrino masses. 
It is possible that the mechanisms responsible for these observations are intrinsically linked.
For example, the observed neutrino masses may arise from  new interactions which at low energies generate the Weinberg operator \cite{Weinberg:1979sa}, ${\cal L}_W = (1/\Lambda)\:\Phi\Phi l_L l_L$, where $\Phi$ is the SM Higgs, $l_L$ are the SM lepton doublets and $\Lambda$ is the scale of new physics. 
${\cal L}_W$ is the lowest-order (non-renormalisable) operator containing only SM fields which can give rise to neutrino masses. Moreover, it violates the lepton-number ($L$) symmetry of the SM, thus suggesting that the interactions which give rise to it may have also generated the BAU via leptogenesis at early times \cite{Fukugita198645}. It is also plausible that the relic abundance of some of the degrees of freedom mediating ${\cal L}_W$ constitutes the DM of the universe today \cite{Dodelson:1993je,Shi:1998km,Kusenko:2006rh,Petraki:2007gq,Kusenko:2010ik,Canetti:2012vf,Canetti:2012kh}.

Another possibility is that the BAU was generated by a mechanism unrelated to the physics of neutrino masses. Well-motivated scenarios include the Affleck-Dine mechanism in supersymmetric models~\cite{Affleck1985361,Dine:1995kz}, and baryogenesis via a first-order phase transition~\cite{Riotto:1999yt}. Clearly, whatever mechanism of baryogenesis operated in the early universe, the existence of a baryon asymmetry today has to be consistent with the interactions which give rise to the neutrino masses. If the latter violate $L$, then they could pose a challenge for the BAU: $L$-violating interactions operating efficiently in the early universe, together with the rapid electroweak (EW) sphaleron processes~\cite{PhysRevD.30.2212,Kuzmin198536}, which violate the baryon-plus-lepton number of the SM, can erase any baryon-number ($B$) asymmetry created before the EW phase transition. It is then important to understand what conditions the survival of a baryon asymmetry imposes on neutrino mass generation mechanisms.

The three minimal, tree level, UV completions of the Weinberg operator, the type-I~\cite{Minkowski:1977sc,Yanagida:1979as,GellMann:1980vs,Mohapatra:1979ia}, type-II~\cite{Magg:1980ut,Schechter:1980gr,Wetterich:1981bx,Lazarides:1980nt,Mohapatra:1980yp,Cheng:1980qt}, and type-III~\cite{Foot:1988aq} seesaw mechanisms, have been studied extensively for their phenomenological consequences. In these scenarios, the scale of new physics ranges from $\sim$100~GeV for Yukawa couplings of $\mathcal{O}(10^{-6})$, all the way up to $\sim 10^{14}$~GeV for Yukawa couplings of $\mathcal{O}(1)$. The small Yukawa couplings in the former case, and the very high energy scale in the latter case make these models difficult to test experimentally (although for the type-II and III seesaws gauge interactions allow the lower mass range to be tested at the LHC).

On the other hand, the inverse seesaw (ISS) mechanism~\cite{PhysRevD.34.1642,Bernabéu1987303}
can generate the observed neutrino masses via new physics at an experimentally accessible scale, and with $\mathcal{O}(1)$ Yukawa couplings. In the ISS, the smallness of neutrino masses is traced to the smallness of an $L$-violating parameter, which we shall denote by $\mu$. This small lepton-number violation naturally suppresses the washout of a pre-existing baryon asymmetry~\cite{Blanchet:2009kk}.
To explain why $\mu$ is small, it is natural to suppose that it is generated radiatively~\cite{Ma:2009gu,Khalil:2010iu,Bazzocchi:2010dt,Law:2012mj,Guo:2012ne}. (Other possibilities of explaining why $\mu$ is a {\em small} parameter are extra dimensions~\cite{Park:2009cm,Fong:2011xh}, or a type-II seesaw-like mechanism~\cite{Dias:2011sq,Dias:2012xp}.)

Radiative ISS models introduce new fields and lepton-number violating interactions, which could potentially wash out the BAU. In this paper, we investigate the constraints on the parameters of the ISS implied by the BAU, or conversely, the conditions on baryogenesis assuming that neutrino masses are generated via the ISS.
We consider two radiative ISS models: one in which lepton number is explicitly broken~\cite{Law:2012mj}, and one in which it is spontaneously broken~\cite{Ma:2009gu}. 
These models also feature accidental discrete symmetries which stabilise exotic particles. The latter can play the role of DM. We investigate the parameter space, derive conditions which ensure that the relic abundance of the new stable particles does not overclose the universe, and identify more specifically the regions which give rise to the observed DM abundance. Similar issues in a different implementation of the ISS have been studied in ref.~\cite{JosseMichaux:2011ba}.

The order of the paper is as follows. In section \ref{sec:iss} we introduce the ISS, and review the natural suppression of washout in the ISS. In section \ref{sec:lawmac} we introduce the Law/McDonald radiative ISS model \cite{Law:2012mj}, analyse washout with the additional fields present, and examine its DM candidates. In section \ref{sec:ma} we repeat the analysis for the Ma radiative ISS model \cite{Ma:2009gu}.

\section{The inverse seesaw mechanism}
\label{sec:iss}
\subsection{The generic mass matrix}
The ISS \cite{PhysRevD.34.1642,Bernabéu1987303} is generically realised with the additional terms in the Lagrangian
\begin{equation}
\mathcal{L} \supset \lambda_{\nu} \epsilon^{ab} \overline{l_{La}} \Phi_{b}^{\dagger} N_{R} + M_{R}\overline{N_{R}}S_{L} + \frac{1}{2}\mu \overline{(S_{L})^{c}}(S_{L})+ \frac{1}{2}\tilde{\mu} \overline{(N_{R})^{c}}(N_{R})+H.c.,
\end{equation}
where $\Phi\sim(1,2,1/2)$ is the usual SM Higgs doublet, $l_{L}\sim(1,2,-1/2)$ is the SM lepton doublet, and $N_R$ and $S_L$ are singlets under the SM gauge group. We leave the details of additional symmetries and symmetry breaking which results in the above terms \cite{Law:2013gma} unspecified for now (a field redefinition ensures the absence of the $\overline{l_{L}} \Phi^{\dagger} (S_L)^c$ term). After $\Phi$ gains a vacuum expectation value (VEV) $\langle\Phi\rangle=(0 \; , \; v_{w}/\sqrt{2})^{T}$ and breaks the EW symmetry $SU(2)_{L}\otimes U(1)_{Y} \to U(1)_{\rm EM}$, the  neutral-fermion mass matrix is (assuming one generation)
\begin{equation}
\label{eq:genericiss}
\mathcal{L} = \frac{1}{2}
\begin{pmatrix}\overline{\nu_{L}} & \overline{N_{R}^{c}} & \bar{S}_L \end{pmatrix}
\begin{pmatrix}
		0 & m_{D} & 0 		\\
		m_{D} & \tilde{\mu} & M_{R}		 \\
		0 & M_{R} & \mu
		\end{pmatrix} 		
\begin{pmatrix}(\nu_{L})^{c} \\ N_{R} \\ (S_{L})^{c} \end{pmatrix},
\end{equation}
where $m_{D} \equiv \lambda_{\nu} v_{w}/\sqrt{2}$. Diagonalization of the mass matrix for $\mu, \tilde{\mu} \ll m_{D},M_{R}$ leads to a light neutrino mass
	\begin{equation}
	m_{\nu} \approx \frac{m_{D}^{2}}{M_{R}^{2}}\mu \ .
	\label{eq:m nu ISS}
	\end{equation}
The heavy Majorana neutrinos $N_{1}$ and $N_{2}$, superpositions of mostly $(S_{L})^{c}$ and $N_{R}$, have mass eigenvalues
	\begin{equation}
	\label{split}
	M_{N} \approx M_{R} \pm \frac{\mu}{2} \ .	
	\end{equation}
Due to the small mass splitting, they form a pseudo-Dirac pair. The mixing between the light and heavy states is approximately $\theta\simeq m_{D}/M_{R}$. In the limit $\mu\to 0$ the light neutrinos become massless at tree level, although loop corrections  give finite contributions to their masses unless both $\mu,\tilde{\mu} \to 0$ \cite{Dev:2012sg}. In this limit one also recovers a global lepton number symmetry in the Lagrangian and hence small choices of $\mu$, and $\tilde{\mu}$ are technically natural. This is the small parameter generated radiatively in the ISS models we return to below.

This discussion can be generalised to a realistic three generation case. The entries in the mass matrix (\ref{eq:genericiss}), are promoted to $3\times3$ matrices, and the light neutrino mass matrix is then to first order,
	\begin{equation}
	M_{\nu} = M_{D}M_{R}^{-1}\mu(M_{R}^{T})^{-1}M_{D}^{T}.
	\end{equation}
Various studies on the collider phenomenology \cite{Dev:2012zg,Das:2012ze}, flavour violating processes \cite{Abada:2007ux,Malinsky:2009df,delAguila:2008pw,Zhang2010297}, and neutrinoless double beta decay \cite{LopezPavon:2012zg,Ibarra:2010xw} of the ISS have appeared in the literature.

\subsection{ISS and baryogenesis}
\label{sec:issbg}
Before the EW phase transition, the non-perturbative sphaleron processes rapidly wash out any existing $B+L$ asymmetry, while they preserve $B-L$. However, if there exist other rapid $L$ violating processes, then both $B$ and $L$ will be driven to zero. In the ISS, the SM lepton number is violated by the parameters $\mu$ and $\tilde{\mu}$. Provided that these parameters are small enough, the $B$ asymmetry is not washed out.

\begin{figure}[h]
\begin{center}
\includegraphics[width=200pt]{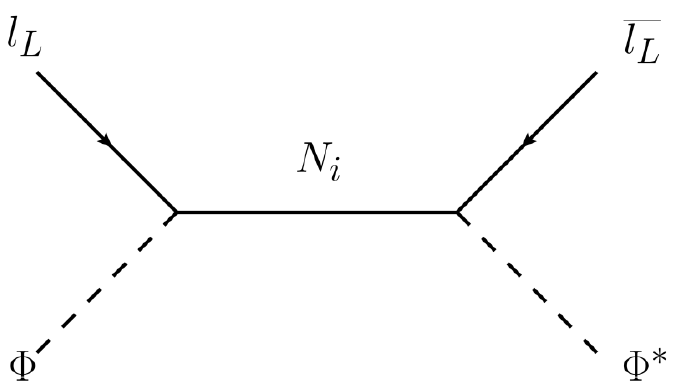}
\end{center}
\caption{$\Delta L=2$ process which can wash out the baryon asymmetry at high temperatures. Both heavy mass eigenstates $N_{1}$ and $N_{2}$ contribute, but destructive interference for small mass splittings $\mu\ll M_N$ crucially limits the washout rate.}
\label{fig:isswashout}
\end{figure}

Let us examine the possible washout of a $B$ asymmetry created at a temperature above the heavy neutrino mass scale, $T_{\rm BG}>M_{N}$, where $T_{\rm BG}$ is the temperature of baryogenesis. In this case, the inverse decays to on-shell heavy Majorana neutrinos $l_{L}+\Phi \to N_{i}$ are not Boltzmann suppressed. Such inverse decays followed by decays $N_{i} \to \overline{l_{L}}+\Phi^{\ast}$ can change the lepton number by two units. This washout process is shown in figure \ref{fig:isswashout}.

It was found in ref. \cite{Blanchet:2009kk} that in the ISS such washout processes can be naturally suppressed, and we review the findings here. For simplicity we take from now on $\tilde{\mu}=0$, as this parameter does not enter the neutrino mass eigenvalues to first order, and focus on $\mu$ instead. The more in-depth discussion of \cite{Blanchet:2009kk} allows both parameters to be non-zero with no major differences to the discussion here. 

For typical mass choices the heavy neutrinos can decay into a component of the Higgs doublet and a lepton $N \to \phi+l_L$. For heavy neutrino masses well above the Higgs mass, $M_{N} \gg M_{h}$, the decay rate is
	\begin{equation}
	\Gamma_{D} = \frac{\lambda_{\nu}^{2}}{8\pi}M_{N},
	\label{eq:heavydecay}
	\end{equation}
and for temperatures $T>M_{N}$ one includes a Lorentz factor $(M_N/T)$. In the type I seesaw, the effectiveness of washout is well captured by the washout parameter~\cite{Kolb:1990vq}
	\begin{equation}
	K \equiv \frac{\Gamma_{D}}{H}\bigg|_{T=M_N},
	\label{eq:washoutparameter}
	\end{equation} 
where $H \sim  g^{1/2}T^{2} / M_{Pl} $ is the expansion rate of the universe, and $g$ is the number of relativistic degrees of freedom. Typically washout processes are ineffective for $K \lesssim 1$, while any net B is erased for $K \gg 1$. In the intermediate regime $K \sim 1-100$ a more careful analysis is required. For our purposes however, we adopt the approximation that $K \lesssim 1$ is safe from washout.

In the ISS, however, the small mass splitting between the heavy pseudo-Dirac neutrinos leads to destructive interference in the scattering process depicted in figure \ref{fig:isswashout}. This leads to a suppressed washout rate \cite{Blanchet:2009kk}
	\begin{equation}
	K_{\rm eff} = K\delta^{2}\equiv\frac{\mu^{2}}{\Gamma_{D} H}\bigg|_{T=M_N},
	\end{equation}
where $\delta \equiv \mu/\Gamma_D$. As expected, in the limit of lepton number conservation, $\mu\to0$, the heavy neutrinos form a Dirac pair, there is complete destructive interference in the $\Delta L=2$ scattering process, and washout of the baryon asymmetry does not occur.  Requiring $K_{\rm eff} \lesssim 1$ to avoid washout translates into a bound:
	\begin{equation}
	\mu < \lambda_{\nu} \left(\frac{M_{N}}{1 \; \mathrm{ \; TeV}}\right)^{3/2}\times6 \; \mathrm{keV}.
	\end{equation}
For $\lambda_{\nu}=0.1$, $M_{N}=1$ TeV, light neutrino masses $m_{\nu}=0.1$ eV are explained by $\mu =330$~eV (cf. eq.~\eqref{eq:m nu ISS}). So the light neutrino masses can be explained while avoiding washout of the BAU. Alternatively, one can express the above bound in terms of the mixing angle:
	\begin{equation}
	\theta\gtrsim 1.4\times10^{-2}\left( \frac{ m_{\nu} }{0.1 \; \text{eV}} \right)^{1/3}\left(\frac{1 \; \mathrm{ \; TeV}}{M_{N}}\right)^{5/6}\left( \frac{ 100 }{g} \right)^{1/6}.
	\end{equation}
To satisfy current experimental constraints (e.g. from universality of the weak interaction and rare leptonic decays) requires roughly, for $M_{N}$ above the EW scale, a mixing angle $\theta \lesssim \mathcal{O}(10^{-2})$ \cite{Antusch:2008tz,Malinsky:2009df}. (The usually stringent constraint on the active-sterile mixing angle from neutrinoless double beta decay \cite{Smirnov:2006bu} is greatly weakened due to the small lepton number violation in the ISS \cite{LopezPavon:2012zg,Ibarra:2010xw}.) Future experimental tests at the $\theta\sim10^{-3}$ level can therefore help test the parameter space corresponding to washout avoidance \cite{Antusch:2009pm}. 

For baryogenesis temperatures below the heavy neutrino scale, $T_{\rm BG}<M_{N}$, the number density of heavy states is Boltzmann suppressed. This means the thermally averaged rate of the net washout process $l_{L}+\Phi \to N_{i} \to \overline{l_{L}}+\Phi^{\ast}$ is suppressed by both a Boltzmann factor and the small $\mu$. In this regime one is even safer from washout. 

What about baryogenesis itself in the ISS? The obvious candidate is resonant leptogenesis \cite{Garny:2011hg}. The resonance itself is due to an almost degenerate pair of heavy neutrinos and exactly such pairs appear in the ISS. Detailed studies have shown resonant leptogenesis is indeed possible in the ISS framework \cite{Blanchet:2010kw}, and it is even possible to link this with the DM abundance \cite{Davidson:2012fn}.

To explain the smallness of $\mu$ we now turn to models where it is generated radiatively. Because of the smallness of $\mu$, washout of the BAU is suppressed in ISS models. However, to generate a finite value of $\mu$ radiatively requires the addition of new fields and $L$-violating interactions. We will see the effects of these new fields and interactions on washout of the BAU, and study the DM candidates in the radiative ISS models. We study the Law/McDonald \cite{Law:2012mj} and Ma radiative ISS models in turn \cite{Ma:2009gu}.

\section{Law/McDonald radiative inverse seesaw}
\label{sec:lawmac}
\subsection{Review of the model}

The Law/McDonald radiative ISS model~\cite{Law:2012mj} employs explicit violation of $(B-L)$, so that processes other than scattering through the heavy neutrinos crucially contribute to washout of the BAU. The model introduces an additional abelian gauge symmetry to the standard model gauge group
	\begin{equation}
	G = G_{\rm SM}\otimes U(1)_{d} \ .
	\end{equation}
All the SM particles are neutral under this new charge. The following exotic fermions are introduced (with convention $Q_{\rm EM}=I_{3}+Y$):
	\begin{equation}
	E_{R,L} \sim (1,3,-1)(0) \quad
	N_{R,L} \sim (1,1,0)(1).
	\end{equation}
Along with the SM Higgs $\Phi\sim(1,2,1/2)(0)$, the following scalars are introduced
	\begin{equation}
	\xi \sim (1,3,-1)(-1),\quad \eta\sim(1,1,0)(1),\quad \chi \sim (1,1,0)(2).
	\end{equation}
The following Yukawa interactions and bare Dirac mass terms are allowed
	\begin{equation}
	\label{eq:lawlagrangian}
	\begin{array}{l l}
	-\mathcal{L} \supset & +y_{E}\overline{l_{L}}\Phi E_{R} + h_{\xi}\overline{E_{L}}\xi N_{R} + h^{'}_{\xi}\overline{E_{R}}\xi N_{L} + h_{\chi}\overline{N_{R}}N_{R}^{c}\chi + h^{'}_{\chi}\overline{N_{L}}N_{L}^{c}\chi \\
	& + M_{E}\overline{E_{L}}E_{R} + M_{N}\overline{N_{L}}N_{R} + H.c.
	\end{array}
	\end{equation}
The VEV pattern is chosen to be
	\begin{equation}
	\langle \phi^{0} \rangle = v_{w}/\sqrt{2}= 174 \; \mathrm{GeV}, \quad 
	\langle \xi^{0} \rangle, \, \langle \eta\rangle=0, 
	\quad \langle \chi \rangle = v_{\chi}/\sqrt{2} \neq 0 \ .
	\end{equation}
The tree level mass matrix for the neutral fermions is then
	\begin{equation}
	\mathcal{L} = \frac{1}{2}
  		\begin{pmatrix}\overline{\nu_{L}} & \overline{(E^{0}_{R})^{c}} & \overline{E^{0}_{L}} \end{pmatrix}\begin{pmatrix}
		0 & \quad y_{E} v_{w}/\sqrt{2} & \quad 0 		\\
		y_{E}v_{w}/\sqrt{2} & 0 & \quad M_{E}		 \\
		0 & M_{E} & \quad 0
		\end{pmatrix} 		\begin{pmatrix}(\nu_{L})^{c} \\ E^{0}_{R} \\ (E^{0}_{L})^{c} \end{pmatrix}.
	\end{equation}
Note $N_{L,R}$ are heavy Majorana fermions decoupled from the above matrix, and we have not yet taken into account the radiative corrections which will generate the terms $\tilde{\mu}\overline{(E^{0}_{R})^{c}}E^{0}_{R}$ and $\mu\overline{(E^{0}_{L})^{c}}E^{0}_{L}$ after symmetry breaking. The scalar potential is given by,
\begin{align}
V_{S} & = (\Phi^\dagger\Phi)\left[\mu_\Phi^2 + \lambda_{\phi}(\Phi^\dagger\Phi)
   + \lambda_{\phi\xi}(\xi^\dagger\xi)
      + \lambda_{\eta \phi}(\eta^\dagger\eta) + \lambda_{\phi\chi}(\chi^\dagger\chi)\right] \nonumber\\
&\;\;+(\xi^\dagger\xi)\left[\mu_\xi^2 + \lambda_{\xi}(\xi^\dagger\xi)
      + \lambda_{\xi\eta}(\eta^\dagger\eta)
         + \lambda_{\xi\chi}(\chi^\dagger\chi)
\right]\nonumber\\
\label{eq:scalarpotential}
&\;\;+(\eta^\dagger\eta)\left[\mu_\eta^2 
      + \lambda_{\eta}(\eta^\dagger\eta)
         + \lambda_{\eta\chi}(\chi^\dagger\chi)
\right] \\
&\;\;
+(\chi^\dagger\chi)\left[\mu_\chi^2 
         + \lambda_{\chi}(\chi^\dagger\chi)
   \right]\nonumber\\
&\;\;
+\left[\lambda_{\xi\phi\eta}\, \xi\Phi\Phi\eta
     +\frac{1}{2}\mu_{\eta\chi}\, \eta\eta\chi^\dagger
     + \text{H.c.}\right]. \nonumber
\end{align}

\begin{figure}[h]
\begin{center}
\includegraphics[width=200pt]{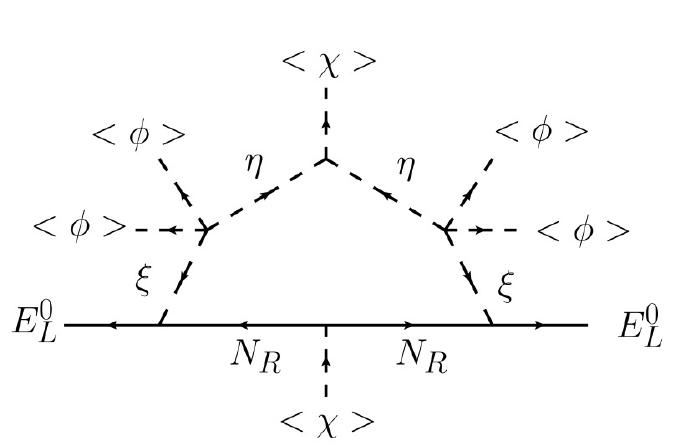}
\end{center}
\caption{Radiative mass generation in the Law/McDonald model.}
\label{fig:law}
\end{figure}

The SM lepton number is violated explicitly by the combination of eq.~\eqref{eq:scalarpotential} and eq.~\eqref{eq:lawlagrangian}, but a Majorana mass term for the $E_{L}$ and $E_{R}$ fields is forbidden due to hypercharge. After $\chi$ and $\Phi$ gain VEVs, a radiative mass term $\mu\overline{(E^{0}_{L})^{c}}E^{0}_{L}$ is generated, as shown in figure \ref{fig:law}. A mass term $\tilde{\mu}\overline{(E^{0}_{R})^{c}}E^{0}_{R}$ is also generated radiatively. The radiative mass is approximately
	\begin{equation}
	\label{eq:radiativemass}
	\mu \approx \frac{1}{16\pi^{2}}\frac{h_{\xi}^{2}h_{\chi}\lambda_{\xi \phi \eta}^{2}}{120}\frac{v_{w}^{4}v_{\chi}^{2}}{\Lambda_{\rm ISS}^{6}}\mu_{\eta \chi},
	\end{equation}
where $\Lambda_{\rm ISS}\sim M_{N},M_{\eta},M_{\xi}$ is the scale of the new physics, usually taken to be $\sim \mathcal{O}(\mathrm{TeV})$, and one may choose $\mu_{\eta \chi}\sim \mathcal{O}(\mathrm{TeV})$. This radiatively generated $\mu$ is exactly the small parameter required in the mass matrix for the ISS. The light neutrinos masses are then
\begin{equation}
\label{eq:examplechoice}
m_{\nu} \approx 
\left(\frac{ y_{E}^{2}h_{\xi}^{2}h_{\chi}\lambda_{\xi \phi \eta}^{2} }{ 10^{-5} }\right) 
\left( \frac{ v_{\chi} }{ M_{E} } \right)^{2} 
\left( \frac{ 1 \text{ TeV} }{ \Lambda_{\rm ISS} } \right)^{6} 
\left( \frac{ \mu_{\eta \chi} }{ 1 \; \text{TeV} } \right) \times 0.1 \text{ eV} \ .
\end{equation}

\subsection{Constraints from BAU washout}
\label{sec:caseblaw}

Previously we have seen that in the ISS framework, the washout of the BAU is naturally suppressed due to the smallness of the lepton-number violating parameter $\mu$. In the Law/McDonald radiative ISS model, $\mu$ is in fact generated only after EW symmetry breaking and cannot itself cause washout. 
However, the SM lepton number is explicitly violated by the couplings which eventually generate $\mu$ and give rise to the radiative ISS. The interactions of eqs. \eqref{eq:scalarpotential} and \eqref{eq:lawlagrangian} can then lead to washout. 
Here, we investigate under what conditions the BAU is not washed out.

\begin{figure}[h]
\begin{center}
\includegraphics[width=350pt]{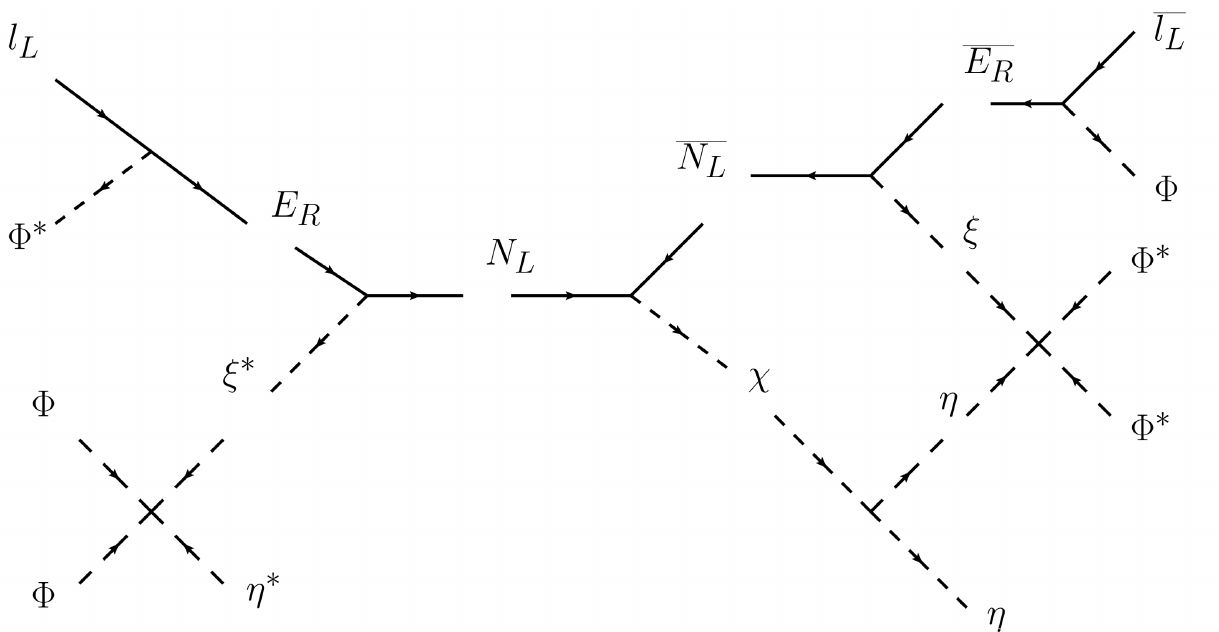}
\end{center}
\caption{An example of a sequence of interactions present in the early universe which shifts the lepton number.}
\label{fig:lawwashout}
\end{figure}

The $(B-L)$ of the SM remains a good symmetry if any of the following sets of parameters vanishes
	\begin{align}
	\label{eq:lrestoration}
	\{y_{E}\}, \;
	\{\mu_{\eta \chi}\}, \;
	\{\lambda_{\xi \phi \eta}\}, \;
	\{h_{\xi}, \; h_{\xi}^{'}\}, \;
	\{h_{\chi}, \; h_{\chi}^{'}\} \; .
	\end{align} 
(Note that $M_E$ has to be large since $E_{R,L}$ are charged under the SM gauge group.)
The BAU can change only if interactions which involve all of the above couplings are efficient. We show such a sequence of interactions in figure \ref{fig:lawwashout}. 
In order to preserve the BAU, we require that the interactions mediated by at least one of these sets of couplings are out of equilibrium in the period between baryogenesis and the EW phase transition. We discern two cases, depending on the scale of baryogenesis:  $T_{\rm BG} > \Lambda_{\rm ISS}$ and  $T_{\rm EW} < T_{\rm BG} < \Lambda_{\rm ISS}$.

\begin{itemize}

\item $T_{\rm BG} > \Lambda_{\rm ISS}$

\begin{enumerate}[(i)]
\item
The Yukawa couplings of eq.~\eqref{eq:lawlagrangian} induce 2-body decays of the heaviest particle participating in each operator, e.g. $N_R \to E_L + \xi^*$. Assuming these decays are kinematically allowed, they proceed with thermally averaged decay rate
\begin{equation}
\label{eq:inversedecaytwo}
  \Gamma_{D} \approx \frac{ h_j^2 \, M_j }{ 16\pi }\left(\frac{ M_j }{T}\right) \ ,
\end{equation}
where $M_j \sim \Lambda_{\rm ISS}$ is the mass of the decaying particle, $h_j$ is the relevant Yukawa coupling, and from now on Lorentz factors, such as $(M_{j}/T)$, appearing in decay rates are understood to be present only for temperatures above the mass of the decaying particle. To prevent washout one requires $(n_{j}/n_{l})\Gamma_{D}\lesssim H \sim  g^{1/2}T^{2} / M_{Pl}$, for $T_{\rm EW} < T < T_{\rm BG}$, where $n_{j}$ $(n_{l})$ is the number density of species $j$ (leptons). The strongest constraint arises from the latest time the density of the exotic particles is still significant: at $T\approx\Lambda_{\rm ISS}$. This implies,
\begin{align}
\label{eq:Yukawa bound}
h_j \lesssim \left( 16\pi g^{1/2} \: \frac{\Lambda_{\rm ISS}^3 }{ M_{Pl} } \right)^{1/2} 
\approx 10^{-7}\left(\frac{ g }{ 100 }\right)^{1/4}
\left(\frac{ \Lambda_{\rm ISS} }{ \text{TeV} } \right)^{1/2} \ , 
\end{align}
where $g$ are the relativistic degrees of freedom at $T\sim M_{j} \sim \Lambda_{\rm ISS}$.

If this condition holds for $y_E$, or $h_\xi$ and $h_\xi'$, or  $h_\chi$ and $h_\chi'$, the lepton-number violating decays and inverse decays induced by these couplings are rare, and the BAU remains frozen.
\item
The thermally averaged rate of the decay mode $\chi \to \eta\eta$ is,
\begin{equation}
\Gamma_{D \; \chi \to \eta \eta} \approx \frac{ \mu_{\eta \chi}^{2} }{ 32\pi\mu_{\chi} }\left( \frac{ \mu_{\chi} }{T}\right).
\end{equation}
Hence the lepton number of the universe is approximately conserved if:
\begin{equation}
\mu_{\eta \chi} \lesssim 
\left(32 \pi g^{1/2} \: \frac{\Lambda_{\rm ISS}^3}{M_{Pl}}
\right)^{1/2}
\approx  \left(\frac{ g }{ 100 }\right)^{1/4}\left(\frac{ \Lambda_{\rm ISS} }{ 1 \;\text{TeV} } \right)^{3/2}10^{-4} \; \text{GeV}.
\label{eq:mu eta chi bound}
\end{equation}

\item
The cross-section for the scattering process $\eta^{\ast}+\xi^{\ast} \to \Phi+\Phi$ is	
\begin{equation}
\sigma \sim \frac{ \lambda_{\xi \phi \eta}^{2} }{  16\pi }\frac{ 1 }{ (T^{2}+\Lambda_{\rm ISS}^{2})  } \ ,
\end{equation}
and the corresponding scattering rate is $\Gamma \sim n(T) \, \sigma$. In the relativistic regime, $\Gamma \propto T$, and hence increases with temperature slower the Hubble parameter. In the $T < \Lambda_{\rm ISS}$ regime, $\Gamma$ becomes Boltzmann suppressed. Thus, for sufficiently small coupling
\begin{equation}
\lambda_{\xi \phi \eta} \lesssim 10^{-7}
\left(\frac{ g }{ 100 }\right)^{1/4}
\left(\frac{ \Lambda_{\rm ISS} }{ 1 \;\text{TeV} } \right)^{1/2} \ ,
\label{eq:lambda bound}
\end{equation}
this process is never in equilibrium.

\item
Lepton-number violating processes may also occur via some of the ISS degrees of freedom propagating off-shell. For example the scattering process $N_{R}+\chi^{\ast}\to \overline{N_{R}} \to \overline{E_{L}}+\xi$ has an approximate cross section
\begin{equation}
\sigma \sim \frac{ (h_{\xi}h_{\chi})^{2} }{ 8\pi }\frac{ T^{2} }{ (T^{2}+M_{N}^{2})^{2} },
\end{equation}
which leads to a bound
\begin{equation}
(h_{\xi}h_{\chi}) \lesssim 10^{-7}
\left(\frac{ g }{ 100 }\right)^{1/4}
\left(\frac{ \Lambda_{\rm ISS} }{ 1 \;\text{TeV} } \right)^{1/2} \ .
\label{eq:off-shell bound}
\end{equation}
\end{enumerate}
\medskip
Satisfying \emph{at least one} of the bounds of eqs.~\eqref{eq:Yukawa bound}, \eqref{eq:mu eta chi bound}, \eqref{eq:lambda bound} and \eqref{eq:off-shell bound} ensures no washout of the BAU, if baryogenesis has occured at temperatures $T_{\rm BG} > \Lambda_{\rm ISS}$. However, as can been seen from eq.~\eqref{eq:examplechoice}, these bounds would imply that the contribution to the active neutrino masses generated via this mechanism is insignificant.

\item $T_{\rm EW} < T_{\rm BG} < \Lambda_{\rm ISS}$

The ISS degrees of freedom remain in thermal and chemical equilibrium even at temperatures much below the ISS scale, due to their gauge, Yukawa and scalar interactions. However, at $T_{\rm BG} < \Lambda_{\rm ISS}$, their density is Boltzmann suppressed. This implies that the exotic leptons carry only a small fraction of the net lepton number of the universe. 

Consider for example again the 2-body decay rate of the ISS degrees of freedom via the Yukawa couplings of eq.~\eqref{eq:lawlagrangian}. The decay rates at $T_{\rm BG} < \Lambda_{\rm ISS}$ are $\Gamma_j \approx h_j^2 M_j/16 \pi$, where as before $M_j$ is the mass of the decaying particle and $h_j$ is the Yukawa coupling that causes the decay. The lepton asymmetry will be approximately conserved if
$\Gamma_j n_j \ll H n_{l}$,
where $n_{l} \sim T^3$ is the lepton-number density, carried mostly by the light relativistic SM leptons, and $n_j \approx (M_j T/2\pi)^{3/2} \exp (-M_j/T)$ is the number density of the non-relativistic exotic leptons.

For Yukawa couplings $h_j \sim {\cal O}(1)$, this yields the constraint
\begin{equation}
%\left( 
\frac{ \Lambda_{\rm ISS} }{ T_{\rm BG} } 
%\right) 
\gtrsim 42 \ ,
\label{eq:boltzmann}
\end{equation}
(where we set $M_j \sim \Lambda_{\rm ISS}$). The interaction considered in the case (ii) gives a similar constraint, while the scattering rates in cases (iii) and (iv) are sufficiently suppressed for:
	\begin{equation}
	\frac{ \Lambda_{\rm ISS} }{ T_{\rm BG} } \gtrsim 25.
	\end{equation}
\end{itemize}
We therefore uncover an important feature of this model: baryogenesis has to occur below the ISS scale. If  it occurs before the EW phase transition, the ISS scale has to be sufficiently high, as seen by eq.~\eqref{eq:boltzmann}. If it occurs after the EW phase transition, no limit on $\Lambda_{\rm ISS}$ applies from considering the survival of the BAU today.
Note that in this model, $\mu$ can only be generated after EW symmetry breaking (see eq.~\eqref{eq:radiativemass}), because the exotic fermions which gain a radiative Majorana mass carry hypercharge. This precludes resonant leptogenesis within this model, as leptogenesis must occur before the sphalerons switch off. It may, however, be possible that another of the ISS fields generates the asymmetry, with washout being suppressed by having the masses of at least one of the other exotic fields at a sufficiently high scale above $T_{\rm BG}$. The construction of such a scenario, ensuring the satisfaction of the Sakharov conditions \cite{Sakharov:1967dj} and generation of the observed BAU, is beyond the scope of this work.

\subsection{Dark Matter}

In the Law/McDonald model the scalars, $\xi$ and $\eta$, and the fermions, $N_{R}$ and $N_{L}$, are odd under an accidental $Z_{2}$ symmetry. The lightest of these is a DM candidate. Assume for now that the lightest state is in the scalar sector (we will briefly discuss the fermionic case below). The neutral components of $\xi$ and $\eta$ mix and the masses of the real and imaginary components split. The general mass squared matrix in the basis $\begin{pmatrix}Re(\eta) & Re(\xi^{0}) & Im(\eta) & Im(\xi^{0})\end{pmatrix}$ is,
	\begin{equation}
	\begin{pmatrix} M_{\eta}^{2}+\mu_{\eta \chi}v_{\chi}/\sqrt{2} & \quad \quad \lambda_{\xi \phi \eta}v_{w}^{2}/4 & 0 & 0 \\ 
			\lambda_{\xi \phi \eta}v_{w}^{2}/4 & M_{\xi}^{2} & 0 & 0 \\
			0 & 0 & \quad  M_{\eta}^{2}-\mu_{\eta \chi}v_{\chi}/\sqrt{2} & \quad \quad -\lambda_{\xi \phi \eta}v_{w}^{2}/4 \\
			0 & 0 & -\lambda_{\xi \phi \eta}v_{w}^{2}/4 & M_{\xi}^{2}
	\end{pmatrix},
	\end{equation}
where we have taken $\lambda_{\xi \phi \eta}$ and $\mu_{\eta \chi}$ to be real and positive without loss of generality as one can absorb the phase by redefining the fields, and
	\begin{align}
	M_{\eta}^{2}\equiv \mu_{\eta}^{2}+\lambda_{\eta \phi}v_{w}^{2}/2+\lambda_{\eta \chi}v_{\chi}^{2}/2, \\
	M_{\xi}^{2}\equiv \mu_{\xi}^{2}+\lambda_{\xi \phi}v_{w}^{2}/2+\lambda_{\xi \chi}v_{\chi}^{2}/2.
	\end{align}	
The lightest of the eigenstates forms the dark matter. If the admixture of $\xi^{0}$ in the DM field is small (which is the case for $M_{\xi} \gg M_{\eta}$) interactions with the $Z$ boson will be suppressed. Scattering of DM with SM particles via the $Z$ or $Z'$ is also suppressed due to the mass splitting between the imaginary and real components of the DM field. (Gauge bosons interact off diagonally with the real and imaginary components of scalar fields.) Annihilation and scattering are dominated by the interactions with the Higgs fields, and for simplicity we will work in this regime. We then have a scalar Higgs portal DM candidate which has been studied extensively in the literature \cite{Silveira:1985rk,McDonald:1993ex,Burgess:2000yq,Cheung:2012xb}. A possible difference from the standard Higgs portal arises when the heavier Higgs boson is light enough to influence the DM annihilation cross section \cite{Aoki:2009pf,Cai:2011kb,He:2011gc,LopezHonorez:2012kv,PhysRevD.86.043511}. We will explore this possibility in greater detail below in the context of the Law/McDonald model, though the discussion is largely applicable to generic Higgs portal models with an EW singlet Higgs mixing with the SM Higgs.

\subsubsection{Double Higgs portal DM}
We denote our (real scalar) DM candidate $\eta$, from now, and take its mass to be:
	\begin{equation}
	M_{DM}^{2}=\mu_{\eta}^{2}+\lambda_{\eta \phi}v_{w}^{2}/2+\lambda_{\eta \chi}v_{\chi}^{2}/2-\mu_{\eta \chi}v_{\chi}/\sqrt{2},
	\label{eq:dmmasssq}
	\end{equation} 
where $\mu_{\eta \chi}>0$. One can choose parameters for the scalar potential (\ref{eq:scalarpotential}) so $\Phi$, and $\chi$ gain the following VEVs: $\langle \Phi \rangle\equiv 1/\sqrt{2} (0 \;,\; v_{w})^{T}$, and $\langle\chi\rangle \equiv v_{\chi}/\sqrt{2}$. The VEVs satisfy the following equations:
	\begin{align}
	\mu_{\chi}^{2}+\frac{1}{2}\lambda_{\phi \chi}v_{w}^{2}+\lambda_{\chi}v_{\chi}^{2}=0, \\
	\mu_{\phi}^{2}+\frac{1}{2}\lambda_{\phi \chi}v_{\chi}^{2}+\lambda_{\phi}v_{w}^{2}=0.
	\end{align}
To ensure that the potential is bounded from below and that $(\Phi,\chi,\eta)=(v_{w}/\sqrt{2},v_{\chi}/\sqrt{2},0)$ is a minimum requires:
	\begin{align}
	&\lambda_{\phi},\lambda_{\chi},\lambda_{\eta}  > 0, \label{eq:bounded}  \\
	&\lambda_{\phi \chi}^{2} < 4\lambda_{\phi}\lambda_{\chi}, \label{eq:nosaddle}  \\
	&M_{DM}^{2}> 0, \label{eq:zerovev} \\	
	&\lambda_{\eta \chi} > -2\sqrt{\lambda_{\eta}\lambda_{\chi}}, \label{eq:boundedtwo}  \\
	&\lambda_{\eta \phi} > -2\sqrt{\lambda_{\eta}\lambda_{\phi}}.  
	\end{align}
An additional condition for symmetry breaking is $\mu_{\chi}^{2}<0$, or $\mu_{\phi}^{2}<0$, and to ensure there is no deeper minimum (at least at zero temperature) one can take $\mu_{\eta}^{2}>0$. The above conditions are necessary but not sufficient to ensure the potential is bounded from below. An additional sufficient condition is:
	\begin{equation}
	\label{eq:sufficient}
	\lambda_{\eta} > \frac{ 1 }{ 4\lambda_{\phi} }\left\{ \frac{ (4\lambda_{\phi}\lambda_{\eta \chi} - 2\lambda_{\phi \chi}\lambda_{ \eta \phi})^{2} }{ 4(4\lambda_{\phi}\lambda_{\chi} - \lambda_{\phi \chi}^{2}) } + \lambda_{\eta \phi}^{2} \right\}.
	\end{equation}
The squared mass matrix for the physical Higgs bosons $\phi$ and $\chi$ is given by:
	\begin{equation}
	\mathcal{L} = 
  		\frac{1}{2}\begin{pmatrix}\phi & \chi \end{pmatrix}\begin{pmatrix}
		2\lambda_{\phi}v_{w}^{2} & \lambda_{\phi \chi}v_{w}v_{\chi} 		\\
		\lambda_{\phi \chi}v_{w}v_{\chi} & 2\lambda_{\chi}v_{\chi}^{2} 	
		\end{pmatrix} 		\begin{pmatrix}\phi \\ \chi \end{pmatrix}.
	\end{equation}
One then finds the mass eigenstates $(m_{H}\geq m_{h})$,
	\begin{align}
	m_{h}^{2}=\lambda_{\chi}v_{\chi}^{2}+\lambda_{\phi}v_{w}^{2}-\sqrt{(\lambda_{\chi}v_{\chi}^{2}-\lambda_{\phi}v_{w}^{2})^{2}+(\lambda_{\phi \chi}v_{w}v_{\chi})^{2}}, \\
	m_{H}^{2}=\lambda_{\chi}v_{\chi}^{2}+\lambda_{\phi}v_{w}^{2}+\sqrt{(\lambda_{\chi}v_{\chi}^{2}-\lambda_{\phi}v_{w}^{2})^{2}+(\lambda_{\phi \chi}v_{w}v_{\chi})^{2}}.
	\end{align}
The mass eigenstates are superpositions of $\phi$ and $\chi$,
	\begin{equation}
	\begin{pmatrix}h \\ H\end{pmatrix}=\begin{pmatrix} \cos{\theta} & \sin{\theta} \\ -\sin{\theta} & \cos{\theta} \end{pmatrix}\begin{pmatrix}\phi \\ \chi\end{pmatrix},
	\end{equation}
where the mixing angle $\theta$ is given by the following,
	\begin{equation}
	\tan{\theta} = \frac{\lambda_{\phi \chi}v_{w}v_{\chi}}{\lambda_{\chi}v_{\chi}^{2}-\lambda_{\phi}v_{w}^{2}+\sqrt{(\lambda_{\chi}v_{\chi}^{2}-\lambda_{\phi}v_{w}^{2})^{2}+(\lambda_{\phi \chi}v_{w}v_{\chi})^{2}}}.
	\end{equation}
Note $h$ ($H$) couples to SM particles the same way as the SM Higgs but with an additional factor of $\cos{\theta}$ ($-\sin{\theta}$) at the vertex. We identify $h$ with the Higgs-like particle discovered at the LHC.

In the limit where the $H$ is very heavy the important DM interactions proceed through the SM-like Higgs, and one in effect returns to the standard Higgs portal scenario. If $H$ is lighter, however, one now has annihilation to SM particles and interactions with quarks proceeding through both $h$ and $H$ propagators. We term this a double Higgs portal and we shall compare the two scenarios below. Similar ideas have been explored previously in the literature: refs. \cite{Aoki:2009pf,Cai:2011kb,He:2011gc} consider two Higgs doublet models, while \cite{LopezHonorez:2012kv} considers fermionic DM with an exotic SM singlet Higgs. We consider scalar DM with an exotic singlet Higgs, a scenario which has been analysed previously in \cite{PhysRevD.86.043511}. One important difference is that we focus more on the phenomenology of high DM masses, while the focus previously has been on the lower mass range $M_{DM} \lesssim 150$ GeV.

In terms of the Higgs mass eigenstates one finds the following interactions with the DM:
	\begin{align}
	-\mathcal{L} \supset 
	& \quad \frac{1}{2}\eta^{2}(\lambda_{\eta \phi}v_{w}\cos{\theta}+ \lambda_{\eta \chi} v_{\chi} \sin{\theta}-\mu_{\eta \chi}\sin{\theta})h \nonumber \\
	& \quad+\frac{1}{2}\eta^{2}(-\lambda_{\eta \phi}v_{w}\sin{\theta} + \lambda_{\eta \chi} v_{\chi} \cos{\theta}-\mu_{\eta \chi}\cos{\theta})H \nonumber \\
	& \quad+\frac{1}{2}\eta^{2}(\lambda_{\eta \chi}\cos{\theta}\sin{\theta} - \lambda_{\eta \phi}\cos{\theta}\sin{\theta})hH  \\
	& \quad+\frac{1}{4}\eta^{2}(\lambda_{\eta \phi}\sin^{2}{\theta}+\lambda_{\eta \chi}\cos^{2}{\theta})H^{2} \nonumber \\
	& \quad+\frac{1}{4}\eta^{2}(\lambda_{\eta \phi}\cos^{2}{\theta}+\lambda_{\eta \chi}\sin^{2}{\theta})h^{2}. \nonumber
	\end{align}
To reduce the number of parameters, and allow one to gain an appreciation of the phenomenology of the double Higgs portal, we study a special case where $\lambda_{\eta \chi} = \lambda_{\eta \phi} \equiv \lambda$ and $\mu_{\eta \chi}=|\lambda| \mu_{\eta \chi}^{'}$, where we will vary $\lambda$ but keep $\mu_{\eta \chi}^{'}>0$ and fixed. The interaction of the Higgs bosons with the DM then simplifies to:
	\begin{align}
	-\mathcal{L} \supset 
	& \quad \frac{1}{2}\Big(v_{w}\cos{\theta}+v_{\chi}\sin{\theta}-\mathrm{Sign}[\lambda] \cdot \mu_{\eta \chi}^{'}\sin{\theta}\Big)\lambda  \eta^{2}h \nonumber \\
	& \quad+\frac{1}{2}\Big(v_{\chi}\cos{\theta}-v_{w}\sin{\theta}-\mathrm{Sign}[\lambda] \cdot \mu_{\eta \chi}^{'}\cos{\theta}\Big)\lambda \eta^{2}H  \label{eq:caseB}\\
	& \quad+\frac{1}{4}\lambda \eta^{2}H^{2}+\frac{1}{4}\lambda \eta^{2}h^{2} . \nonumber
	\end{align}

To see the effect of the second Higgs boson, we present two cases. Case A is where the heavy Higgs boson is taken to be heavy enough and the mixing negligible, so it is effectively decoupled. In this limit one obtains the standard Higgs portal. For case B, the double Higgs portal, we choose some parameters, namely $M_{H}=500$ GeV, $v_{\chi}=1$ TeV, $\mu_{\eta \chi}^{'}=2$ TeV, and examine the change in phenomenology for different choices of the Higgs mixing parameter $\sin{\theta}$.

Furthermore, given the above choice of parameters, we have calculated $\lambda_{\chi}$, $\lambda_{\phi}$, and $\lambda_{\phi \chi}$ as functions of $\sin{\theta}$. One finds that condition (\ref{eq:nosaddle}) is satisfied for the entire range of the mixing angle ($-\pi/2 \leq\theta \leq \pi/2$), and that $\mu_{\chi}^{2} < 0$ so that symmetry breaking can proceed.

As the heavy Higgs boson enters loop corrections to the $W$ and $Z$ boson propagators, a constraint on $\sin{\theta}$ comes from the oblique parameters: $T$, $S$, and $U$. By following the analysis of \cite{Profumo:2007wc,PhysRevD.77.035005,PhysRevD.86.043511} one finds that for $M_{H}=500$ GeV the mixing angle is constrained to be
	\begin{equation}
	\label{eq:ewbound}
 	|\sin{\theta}| \leq 0.43
	\end{equation}
by EW precision observations at 95\% C.L. (assuming the other exotic particles give only negligible contributions.)

\subsubsection{DM relic abundance}
\begin{figure}[h]
	\begin{center}
	\includegraphics[width=400pt]{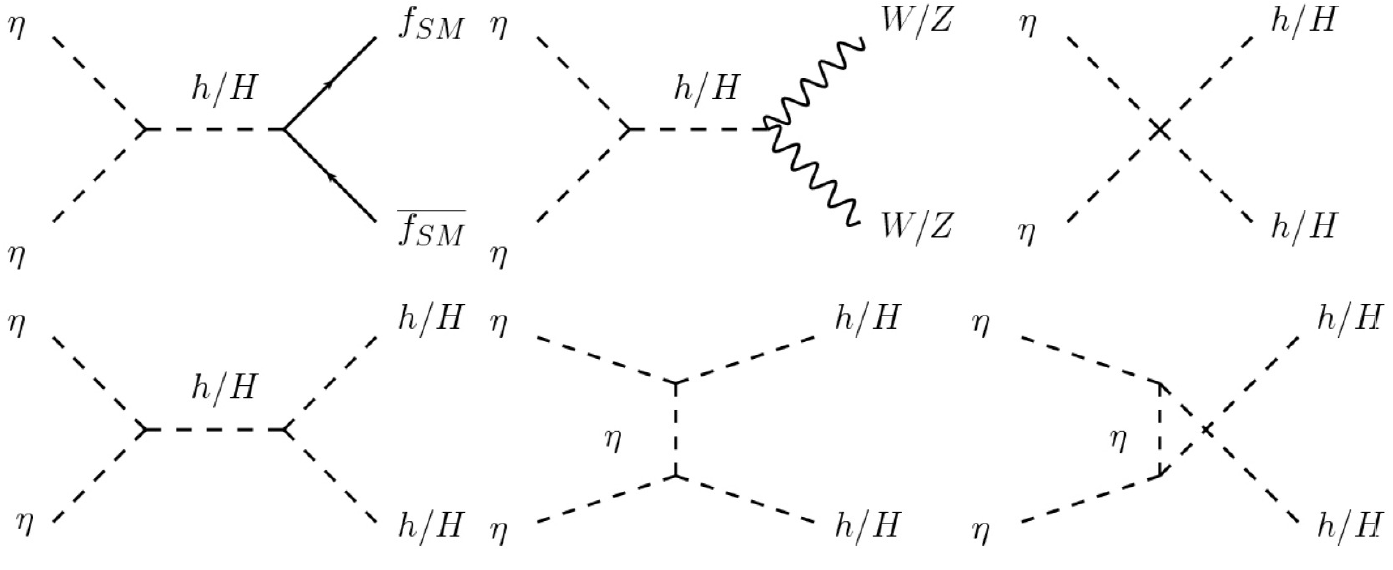}
	\end{center}
  	\caption{Possible DM annihilation channels. The diagrams on the lower line give only a small correction and are not included in the calculations.}
	\label{fig:annihilations}
	\end{figure}

One can then calculate the annihilation cross section of DM into standard model particles. The relevant Feynman diagrams are shown in figure \ref{fig:annihilations}. To obtain the velocity averaged cross section, one should integrate over the velocity distribution. However a good approximation is obtained with,
	\begin{equation}
	\langle\sigma v_{rel}\rangle \simeq (\sigma v_{rel})|_{s=4M_{DM}^{2}}.
	\end{equation}
One can then simply compare to the required annihilation cross section, for the observed relic abundance \cite{Komatsu:2010fb}, which for DM masses greater than 10 GeV is approximately \cite{Steigman:2012nb},
	\begin{equation}
	\langle\sigma v_{rel}\rangle_{relic} \approx 2.2\times 10^{-26} \; \mathrm{cm}^{3}/\mathrm{s}.
	\end{equation}

To obtain the desired $M_{DM}^{2}$ with a choice $\mu_{\eta}^{2}>0$ one wants to choose the required values of $\lambda$ to be negative (particularly for small DM masses), as can be seen from eq. (\ref{eq:dmmasssq}). Some degree of fine tuning is of course required for small DM masses given our choice $v_{\chi}=1$ TeV.

\begin{figure}[h]
	\begin{center}
	\includegraphics[width=450pt]{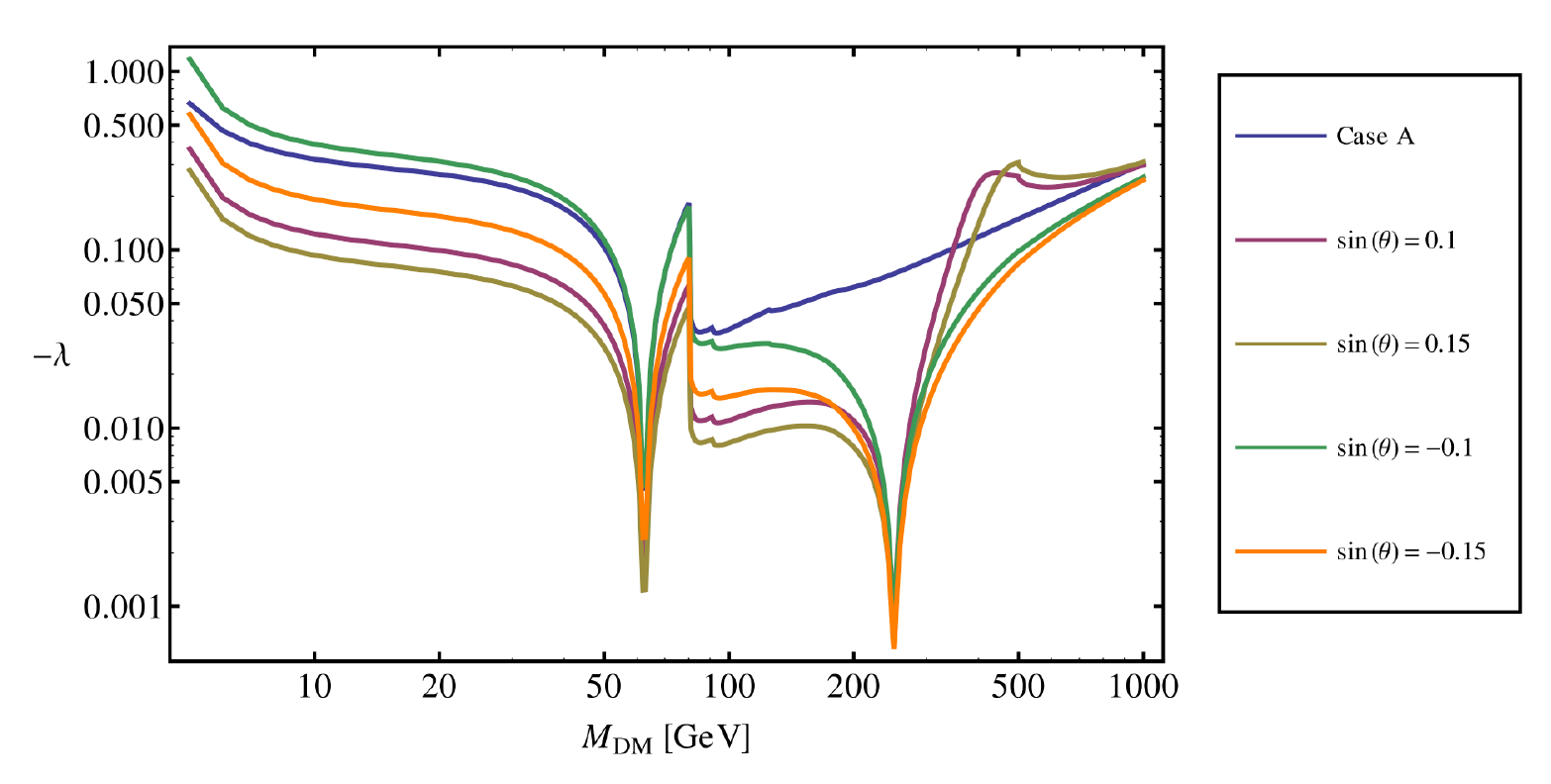}
	\end{center}
  	\caption{Required coupling for the correct DM abundance for different values of $\sin{\theta}$ and Case A (single Higgs portal) for comparison. Parameters have been chosen so that $\lambda \equiv \lambda_{\eta \phi} = \lambda_{\eta \chi}$, $\mu_{\eta \chi}^{'} = 2$ TeV, $M_{H}=500$ GeV, and $v_{\chi}=1$ TeV.}
	\label{fig:relicpositive}
	\end{figure}

We have calculated the required coupling $\lambda \equiv \lambda_{\eta \phi} = \lambda_{\eta \chi}$, with the constraint $\lambda<0$. (The relevant expressions for the cross sections can be found in the appendix of ref. \cite{Cheung:2012xb} for case A, and simple modifications for case B can be found by taking into account the Lagrangian in eq. (\ref{eq:caseB}).) The results for different choices of the mixing parameter are shown in figure \ref{fig:relicpositive}. The observed relic abundance can be achieved with values of $\lambda\sim-0.1$ for most of the DM mass range. Note the falls in the required coupling at the points where $2M_{DM} = M_{h/H}$, as the intermediate Higgs boson is on-shell.

We have also checked that the constraints of eqs. (\ref{eq:boundedtwo}-\ref{eq:sufficient}) can be satisfied with perturbative choices of the quartic coupling $\lambda_{\eta}$ (except for small regions of parameter space around $\sin{\theta}\approx0.05$, where for low DM masses $|\lambda|>1$).

\subsubsection{Direct detection}

	\begin{figure}[h]
	\begin{center}
	\includegraphics[width=150pt]{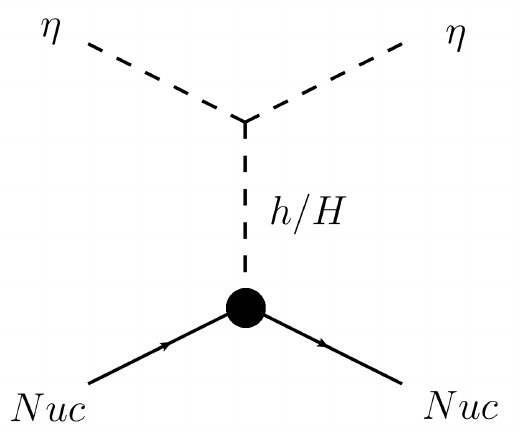}
	\end{center}
  	\caption{DM Nucleon scattering through t-channel Higgs exchange. The effective coupling of the Higgs with the Nucleon is indicated.}
	\label{fig:etan}
	\end{figure}

DM particles will scatter off nuclei through t-channel Higgs boson exchange (see figure \ref{fig:etan}). The spin independent DM-nucleon scattering cross section for our DM candidate $\eta$ is given by,
	\begin{equation}
	\sigma^{SI}_{DM-Nuc} = \frac{\lambda^{2}M_{Nuc}^{4}f_{N}^{2}}{4\pi(M_{DM}+M_{Nuc})^{2}}\left(\frac{A}{m_{H}^{2}}+\frac{B}{m_{h}^{2}}\right)^{2},
	\end{equation}
where $M_{Nuc}\simeq0.94$ GeV is the nucleon mass, $f_{N}$ parametrizes the effective Higgs-nucleon coupling, and the vertex factors are:
	\begin{align}
	A=-\sin{\theta}\Big\{\frac{v_{\chi}}{v_{w}}\cos{\theta}-\sin{\theta}-\mathrm{Sign}[\lambda]\cdot2\frac{ \mu_{\eta \chi}^{'} }{ v_{w} }\cos{\theta}\Big\},  \\
	B=\cos{\theta}\Big\{\cos{\theta}+\frac{v_{\chi}}{v_{w}}\sin{\theta}-\mathrm{Sign}[\lambda]\cdot2\frac{ \mu_{\eta \chi}^{'} }{ v_{w} }\sin{\theta}\Big\}.
	\end{align}
Note that uncertainties in $f_{N}$ result in roughly a factor of $\sim5$ uncertainty in $\sigma^{SI}_{DM-Nuc}$ \cite{PhysRevD.84.115017}. We use a value of $f_N=0.3$ \cite{Toussaint:2009pz,Young:2009ps,Alarcon:2011zs,Alarcon:2012nr}.

\begin{figure}[h]
	\begin{center}
	\includegraphics[width=450pt]{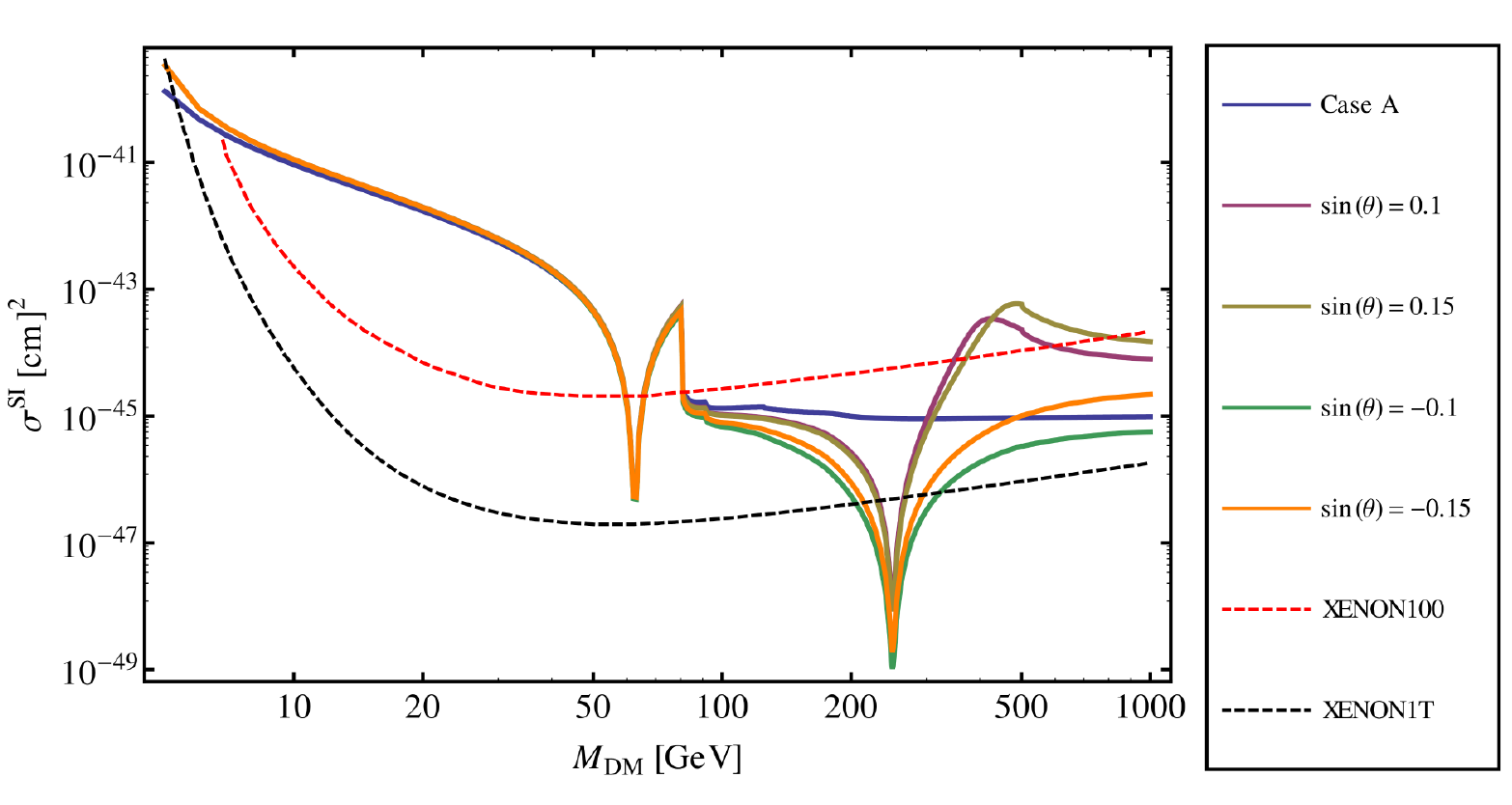}
	\end{center}
  	\caption{DM-Nucleon scattering cross section for case A and case B with different values of sin($\theta$), the upper limit from XENON100, and the projected sensitivity for XENON1T in 2017.}
	\label{fig:directsi}
	\end{figure}

Having calculated $\lambda$, we then find $\sigma^{SI}_{DM-Nuc}$ for different values of $\sin{\theta}$. The results are presented in figure \ref{fig:directsi}. XENON100 exclusion limits from 255 live days of data taking \cite{Aprile:2012nq}, labeled XENON100, and projected XENON1T exclusions for 2017 in the case of no WIMP-nucleon scattering \cite{Aprile:2012zx}, labeled XENON1T, are plotted for comparison. We see that $M_{DM}\gtrsim80$ GeV is still largely viable, but that much of the remaining parameter space can be ruled out by XENON1T. 

Note that due to different interferences the scattering rate can be above or below the zero mixing case. A minimum is reached for $\sin{\theta}=-0.052$, which is exactly where,
	\begin{equation}
	\left(\frac{A}{m_{H}^{2}}+\frac{B}{m_{h}^{2}}\right)=0,
	\end{equation}
so that the DM-nucleon cross section is zero. Similar interference was noted in the case of a two Higgs doublet extension of the Higgs portal in ref. \cite{Cai:2011kb}. 

\subsubsection{Higgs signals at colliders}
The double Higgs portal can be tested in a number of ways at colliders \cite{He:2011gc,Cai:2011kb,Bandyopadhyay:2012px,Englert:2011aa}, some of them common with the usual single Higgs portal \cite{Djouadi:2011aa,PhysRevD.84.115017,Djouadi:2012zc,PhysRevD.77.035005}. Mixing with the exotic Higgs will reduce the production of the SM-like Higgs,
	\begin{equation}
	\sigma_{pp \to h} = \mathrm{cos}^{2}\theta\cdot \sigma_{pp \to h}^{SM}.
	\end{equation}
The total width will also be reduced by a factor of $\mathrm{cos}^{2}\theta$, however the SM-width for $m_{h}=125$ GeV, roughly $4$ MeV, is already far below current experimental resolution. The discovery of a SM-like Higgs boson at the LHC \cite{:2012gu,:2012gk} allows us to constrain this scenario. The observed signal strength over the SM expectation and $1\sigma$ uncertainty is \cite{CMS-PAS-HIG-12-045,ATLAS-CONF-2013-034}:
	\begin{align}
	\sigma/\sigma_{SM} & =0.88 \pm 0.21 \quad \mathrm{(CMS)}, \\
	\sigma/\sigma_{SM} & =1.3 \pm 0.13 \; \text{(stat.)} \pm 0.14 \; \text{(syst.)} \quad \mathrm{(ATLAS)}.
	\end{align}
Combinations of the results put a limit on the universal suppression of all channels so that: $\sigma/\sigma_{SM}\gtrsim0.8$ at the $2\sigma$ level \cite{Ellis:2013lra,Falkowski:2013dza,Giardino:2013bma}. This translates into a limit,
	\begin{equation}
	\label{eq:lhclimit}
	|\sin{\theta}| \leq 0.45,
	\end{equation} 
if one assumes no invisible decays of the 125 GeV resonance. (An alternative, more conservative, treatment of the theoretical uncertainty in $\sigma_{SM}$ modifies the above bound to be $|\sin{\theta}| \lesssim0.7$ instead \cite{Djouadi:2013qya}.) Equation (\ref{eq:lhclimit}) is comparable to the constraint from EW precision observables, eq. (\ref{eq:ewbound}), however the limit on the mixing angle from measurement of the signal strength is generally applicable and not limited to our choice $M_{H}=500$ GeV. 

If the DM is light enough, the Higgs can decay invisibly to two DM particles. The invisible partial decay width of the Higgs $(h \to \eta \eta)$ is given by,
	\begin{equation}
	\Gamma = \frac{	\lambda^{2} }{ 32\pi m_{h} }\left(v_{w}\cos{\theta}+v_{\chi}\sin{\theta}-\mathrm{Sign}[\lambda]\cdot 2 \mu_{\eta \chi}^{'}\sin{\theta}\right)^{2}\left(1-\frac{ 4m_{\eta}^{2} }{ m_{h}^{2} } \right)^{1/2}.
	\end{equation} 
This leads to a large invisible branching fraction $Br(h\to\eta \eta)\gtrsim0.99$ for light DM masses $M_{DM}\lesssim10$ GeV, except for a very narrow region around a mixing angle $\sin{\theta}=-0.049$ where the effective coupling of the $h\eta\eta$ term is zero. This helps to rule out such light DM candidates as the direct detection experiments lose sensitivity for low mass DM candidates. 

If the mixing angle is large enough, the heavy state could also be detected. For a Higgs mass of 500 GeV, CMS data currently sets a 95\% C.L. \cite{Chatrchyan:2013yoa}:
	\begin{equation}
	\sigma/\sigma_{SM}\simeq0.21.
	\end{equation}
The most stringent limit on the mixing angle will come from assuming $2M_{DM} > M_{H}$, so there is no invisible width. There is also an additional complication of $H\to hh$ decays \cite{Bowen:2007ia}, however, for the parameters we have chosen the relevant branching ratio turns out to be negligible. The limit on the mixing angle from high mass Higgs searches is then:
	\begin{equation}
	|\sin{\theta}| \leq 0.46.
	\end{equation}
Thus the most stringent limit on the mixing angle in case B with $M_{H}=500$ GeV comes from considering EW precision observables, eq. (\ref{eq:ewbound}), while the most widely applicable constraint comes from measurements of the signal strength of the 125 GeV resonance, eq. (\ref{eq:lhclimit}), as no assumptions about the mass of the heavier Higgs or VEV have to be taken as an input.

\subsubsection{Fermionic DM}
Alternatively one may consider fermionic Higgs portal DM. The important terms are then:
	\begin{equation}
	-\mathcal{L} \supset h_{\chi}\chi\overline{N_{R}}(N_{R})^{c} + h^{'}_{\chi}\chi\overline{N_{L}}(N_{L})^{c} + M_{N}\overline{N_{L}}N_{R} + H.c.,
	\end{equation}
where the lightest mass eigenstate of the $N_{L},\; (N_{R})^{c}$ admixture is the DM candidate (protected by the same accidental $Z_{2}$ as in the scalar DM case). In such a scenario DM is constrained to be very heavy from direct detection experiments, $M_{DM}\gtrsim 2$ TeV, unless the DM mass is at one of the two Higgs resonances, there is parity violation, or a Sommerfeld enhancement \cite{LopezHonorez:2012kv}. So while we have studied the scalar DM case above, either scalar or fermionic DM is viable in the Law/McDonald model. Further generalizations taking into account interactions with the $Z'$ in the fermionic DM scenario are also possible, but we will not pursue the details here.  

\subsection{Concluding remarks on the Law/McDonald model}
We have seen that in the Law/McDonald model, baryogenesis is severely constrained. The explicit lepton number violation will tend to wash out any baryon asymmetry created at a high scale. However at temperatures roughly a factor of 25-40 below the ISS mass scale, Boltzmann suppression of the thermally averaged reaction rates means that the ISS fields will not contribute to washout (though not necessarily \emph{all} of the ISS fields must have such high masses). Unfortunately the small $L$ violating parameter is only generated after EW symmetry breaking, meaning that resonant leptogenesis is not possible in the Law/McDonald radiative ISS.

We found that the Law/McDonald model can easily accommodate the observed DM abundance through the Higgs portal. Also, in a simple extension, an additional Higgs state can also contribute to the DM annihilation rate and DM nucleon scattering rate. Apart from narrow regions of parameter space, either on resonance points for the annihilation rate, or near complete destructive interference for the DM nucleon scattering rates, these models can for the most part be tested by the XENON1T experiment.

\section{Ma radiative inverse seesaw model}
\label{sec:ma}
\subsection{Review of the model}
We now consider the Ma radiative ISS model \cite{Ma:2009gu}. An important difference in this model is that $(B-L)$ is a protected symmetry down to the ISS scale. First let us review the main features of this model. It employs an additional $U(1)$ which originates from a grand unified theory (GUT) with the following breaking pattern,
	\begin{equation}
	\begin{array}{l l}
	SO(10) & \rightarrow SU(5)\otimes U(1)_{\chi} \\
		& \rightarrow SU(3)_{C}\otimes SU(2)_{L} \otimes U(1)_{Y} \otimes U(1)_{\chi}.
	\end{array}
	\end{equation}
The exotic charge is given by the following,
	\begin{equation}
	Q_{\chi} = 5(B-L)-4Y=5(B-L)+4I_{3L}-4Q_{EM}.
	\end{equation}
There are two scalars, $\eta_{1}$ and $\eta_{2}$, that transform under $U(1)_{\chi}$ as per
	\begin{equation}
	Q_{\chi}(\eta_{1})= 1, \quad Q_{\chi}(\eta_{2}) = 2,
	\end{equation}
and are neutral with respect to the SM gauge group. These will be used for the radiative ISS and to break the new $U(1)_{\chi}$. To allow the usual Yukawa interactions,
	\begin{equation}
	\mathcal{L} = \lambda_{d} \overline{Q_{L}} \cdot \Phi  d_{R} + \lambda_{u} \epsilon^{ab} \overline{Q_{La}} \Phi_{b}^{\dagger} u_{R} +
		      \lambda_{e} \overline{l_{L}} \cdot \Phi  e_{R} + \lambda_{\nu} \epsilon^{ab} \overline{l_{La}} \Phi_{b}^{\dagger} N_{R} + H.c.,
	\end{equation}
the SM Higgs doublet must carry a $U(1)_{\chi}$ charge,
	\begin{equation}
	\Phi = (\phi^{+},\phi^{0})\sim(1,2,1/2,-2).
	\end{equation}
The fermion multiplets of $SO(10)$ contain a right handed neutrino, $N_{R} \sim (1,1,0,-5)$, per generation of SM fermions. To these are added more SM neutral fermions with $Q_{\chi}$ given by,
	\begin{equation}
	1\times S_{3L} \sim -3, \quad 4\times S_{2L} \sim 2, \quad 5\times S_{1L} \sim -1,
	\end{equation}
where the required number of copies has been indicated to form an anomaly-free set. The following Yukawa terms involving the exotic scalars are permitted,
	\begin{equation}
	\label{eq:exoticyukawas}
	\mathcal{L} \supset f_{R3}\overline{N_{R}}S_{3L}\eta_{2}^{\dagger} + f_{23} \overline{(S_{3L})^{c}}S_{2L}\eta_{1}
			   + f_{12}\overline{(S_{2L})^{c}}S_{1L}\eta_{1}^{\dagger} + f_{11}\overline{(S_{1L})^{c}}S_{1L}\eta_{2} + H.c.
	\end{equation}
The scalar potential is given by,
	\begin{align}
	\label{eq:morescalars}
	V = & \; (A\eta_{1}^{2}\eta_{2}^{\dagger} + H.c.) + \lambda_{11} (\eta_{1}^{\dagger}\eta_{1})^{2} + \lambda_{22} (\eta_{2}^{\dagger}\eta_{2})^{2} + \lambda_{12} (\eta_{1}^{\dagger}\eta_{1})(\eta_{2}^{\dagger}\eta_{2}) \nonumber \\
	&+\mu_{\Phi}^{2}\Phi^{\dagger}\Phi + \lambda_{\Phi}(\Phi^{\dagger}\Phi)^{2}+\lambda_{ 1 \Phi}(\Phi^{\dagger}\Phi)(\eta_{1}^{\dagger}\eta_{1})+\lambda_{ 2 \Phi}(\Phi^{\dagger}\Phi)(\eta_{2}^{\dagger}\eta_{2})   \\
	&+\mu_{1}^{2}(\eta_{1}^{\dagger}\eta_{1})+\mu_{2}^{2}(\eta_{2}^{\dagger}\eta_{2}), \nonumber	
	\end{align}
where $A$ is a coupling constant with dimension of mass. The $U(1)_{\chi}$ symmetry is spontaneously broken by,
	\begin{equation}
	\langle\eta_{2}\rangle \equiv v_{\chi}/\sqrt{2} \neq 0,\quad \langle\eta_{1}\rangle = 0.
	\end{equation}
Before radiative corrections are taken into account the neutral fermion mass matrix is given by,
		\begin{equation}
	\label{eq:maissmassmatrix}
	\mathcal{L} = \frac{1}{2}
  		\begin{pmatrix}\overline{\nu_{L}} & \overline{(N_{R})^{c}} & \overline{S_{3L}} & \overline{S_{1L}} \end{pmatrix}\begin{pmatrix}
		0 & m_{D} & 0 & 0		\\
		m_{D} & 0 & M_{N} & 0		 \\
		0 & M_{N} & 0 & 0		\\
		0 & 0 & 0 & M_{1} 		\\
		\end{pmatrix} 		\begin{pmatrix}(\nu_{L})^{c} \\ N_{R} \\ (S_{3L})^{c} \\ (S_{1L})^{c} \end{pmatrix},
	\end{equation}
where $m_{D}=\lambda_{\nu} v_{w}/\sqrt{2}$, $M_{N}=f_{R3} v_{\chi}/\sqrt{2}$, and $M_{1} = \sqrt{2}f_{11}v_{\chi}$. Note the $S_{2L}$ are massless at tree level, and that the scalar interaction term,
	\begin{equation}
	A\eta_{1}\eta_{1}\eta_{2}^{\dagger} + A^{\ast}\eta_{1}^{\dagger}\eta_{1}^{\dagger}\eta_{2} \rightarrow A\eta_{1}\eta_{1}v_{\chi}/\sqrt{2} + A^{\ast}\eta_{1}^{\dagger}\eta_{1}^{\dagger}v_{\chi}/\sqrt{2},
	\end{equation}
induces a mass splitting between the mass of the real ($m_{R}$) and mass of the imaginary ($m_{I}$) components of $\eta_{1}$,
	\begin{equation}
	m_{R}^{2}-m_{I}^{2} = 2\sqrt{2}Av_{\chi},
	\end{equation}
where we set $A$ to be real and positive without loss of generality as any complex phase can be absorbed into the fields.

\begin{figure}[h]
\begin{center}
\includegraphics[width=400pt]{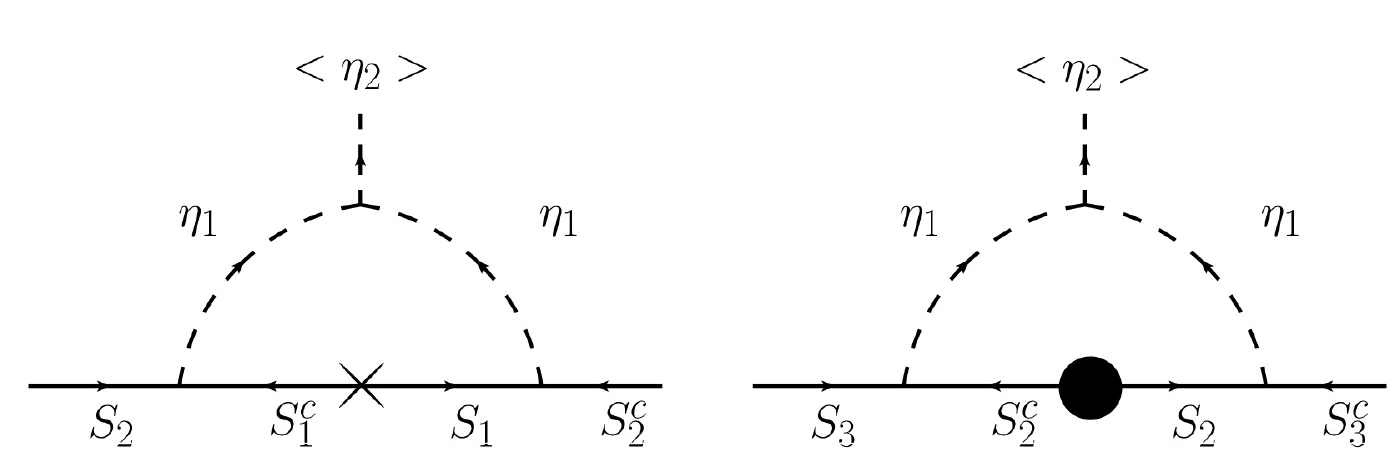}
\end{center}
\caption{Radiative mass generation for $S_{2}$ and $S_{3}$ in the Ma model.}
\label{fig:ma}
\end{figure}

The exchange of $Re(\eta_{1})$ and $Im(\eta_{1})$ then give one-loop radiative Majorana masses to the $S_{2L}$ as can be seen in figure \ref{fig:ma}:
	\begin{equation}
	\label{eq:masstwo}
	M_{2} = \frac{(f_{12})^{2}M_{1}}{16\pi^{2}} 
	\left[\frac{m_{R}^{2}}{m_{R}^{2}-M_{1}^{2}}\mathrm{ln}\left(\frac{m_{R}^{2}}{M_{1}^{2}}\right) - \frac{m_{I}^{2}}{m_{I}^{2}-M_{1}^{2}}\mathrm{ln}\left(\frac{m_{I}^{2}}{M_{1}^{2}}\right) \right].
	\end{equation}
Similarly one obtains two-loop radiative Majorana masses for the $S_{3L}$,
	\begin{equation}
	\label{eq:massthree}
	\mu \equiv M_{3} = \frac{(f_{23})^{2}M_{2}}{16\pi^{2}} 
	\left[\frac{m_{R}^{2}}{m_{R}^{2}-M_{2}^{2}}\mathrm{ln}\left(\frac{m_{R}^{2}}{M_{2}^{2}}\right) - \frac{m_{I}^{2}}{m_{I}^{2}-M_{2}^{2}}\mathrm{ln}\left(\frac{m_{I}^{2}}{M_{2}^{2}}\right) \right],
	\end{equation}
which are exactly the small parameters required in the mass matrix of eq. (\ref{eq:maissmassmatrix}) for the ISS.

Note $\eta_{1}$ and $S_{2L}$ are odd under an accidental $Z_{2}$ symmetry, while all the other particles are even. We will see below, that for typical choices of the parameters, $S_{2L}$ is lighter than $\eta_{1}$ as its mass arises radiatively and therefore $S_{2L}$ is the DM candidate. One can find various approximate forms for the radiative masses depending on the sizes of various parameters. If we take $m_{R}^{2} \gg m_{R}^{2}-m_{I}^{2} = 2\sqrt{2}Av_{\chi}, M_{1}$, eq. (\ref{eq:masstwo}) can be approximated as,
	\begin{equation}
	M_{2} \approx \frac{\sqrt{2}(f_{12})^{2}Av_{\chi}M_{1}}{8\pi^{2}m_{I}^{2}},
	\end{equation}
while for natural choices of the parameters $m_{R}^{2},m_{I}^{2} \gg M_{2}^{2}$, one obtains from eq. (\ref{eq:massthree}):
	\begin{equation}
	\label{eq:massthreeapprox}
	\mu \equiv M_{3} \approx \frac{(f_{12})^{2}(f_{23})^{2}A^{2}v_{\chi}^{2}M_{1}}{32\pi^{4}m_{I}^{4}}.
	\end{equation}
To give a numerical example, for $v_{\chi}\sim M_{1} \sim A \sim \mathcal{O}(10 \; \mathrm{TeV})$, and $m_{I} = 50$ TeV, one finds $M_{2} \approx (f_{12})^{2}\times7$ GeV and $M_{3} \approx (f_{23})^{2}(f_{12})^{2}\times5$ MeV (in very close agreement to what the exact formulas give).

\subsection{Constraints from BAU washout}
\begin{figure}[h]
\begin{center}
\includegraphics[width=400pt]{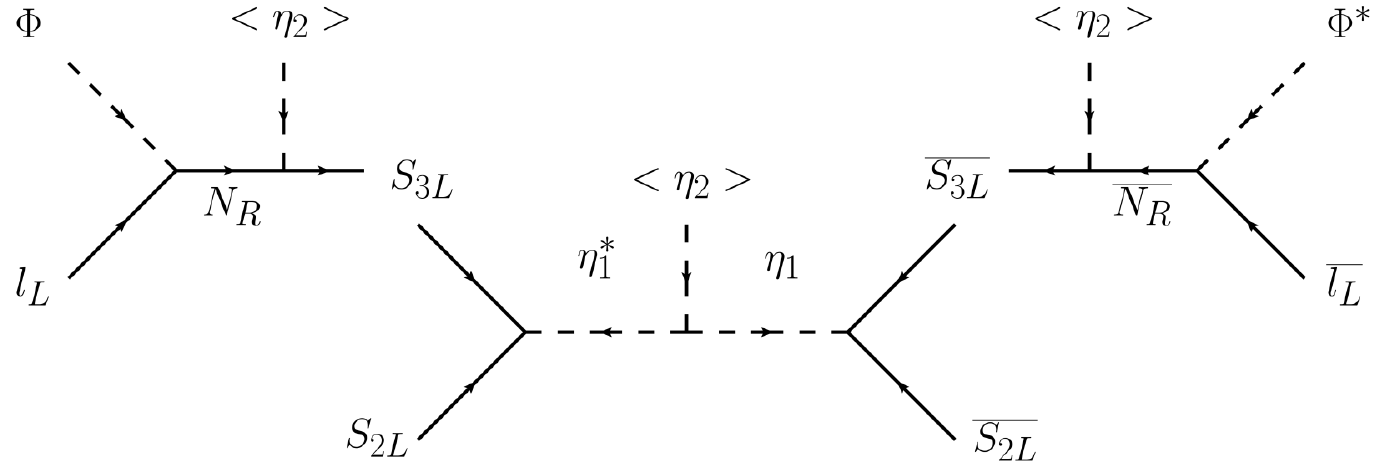}
\end{center}
\caption{Washout process involving the exotic fermions and scalars. Also shown are the Yukawa interactions which link this washout process to the SM sector. Exchange of the real and imaginary components of $\eta_{1}$ leads to destructive interference. Not shown are the interactions involving $f_{12}$ and $f_{11}$ which ensure $S_{2L}$ cannot be assigned non-zero SM lepton number.}
\label{fig:crazywashout}
\end{figure}

Until $U(1)_{\chi}$ is broken, $(B-L)$ is an exact symmetry of the Ma model, and generation or washout of the BAU will not take place.\footnote{This constraint could be circumvented through sequestration of positive and negative $B-L$ charge into the visible sector and a hidden sector. This is a popular mechanism in asymmetric DM models, which aim to dynamically relate the visible and dark matter relic abundances \cite{Nussinov:1985xr,Davoudiasl:2010am,vonHarling:2012yn,Bell:2011tn,Cheung:2011if,PhysRevLett.64.340,Kuzmin:1996he,Kitano:2004sv,Gu:2010ft,Gu:2007cw,Heckman:2011sw,Farrar:2005zd,Petraki:2011mv}. In the context of the model studied here, it would require introducing additional fields and interactions, such that the $B-L$ symmetry encompasses two low-energy global U(1) symmetries -- the $(B-L)_v$ carried by the SM fields and a dark baryon number $B_d$. The construction of such a model in an $SO(10)$ GUT is beyond the scope of this paper.}
The VEV $v_{\chi}$ breaks $Q_{\chi}=5(B-L)-4Y$, and hence the $(B-L)$ as defined above. Baryogenesis can then take place, either through resonant leptogenesis, or some other suitable baryogenesis scenario with $T_{\rm BG}\leq T_{\chi} \sim v_{\chi}$. However, as shown in figure \ref{fig:crazywashout}, the ISS fields can now also wash out the asymmetry. 

A new conserved global $(B-L)$ can be defined, even with $\langle\eta_{2}\rangle\neq0$, provided any of the following parameters vanishes:
	\begin{equation}
	\label{eq:lrestorationma}
	\lambda_{\nu}, \; f_{R3}, \; f_{23}, \; f_{12}, \; f_{11}, \; A.
	\end{equation}
If the interactions mediated by one of the above parameters are out of equilibrium the BAU is safe. We first examine what values the parameters must take so as to not wash out the BAU. In section (\ref{sec:resonantleptogen}) we examine the possibility for resonant leptogenesis, where the lightest pseudo-Dirac pair generates the BAU.

\begin{itemize}
\item $T_{\rm BG} \approx \Lambda_{ISS} \approx v_{\chi}$

\begin{enumerate}[(i)]

\item The decays and inverse decays $N_{R} \leftrightarrow l_{L}+\Phi$ are controlled by $\lambda_{\nu}$. The decay rate is approximately,
	\begin{equation}
	\Gamma \approx \frac{ \lambda_{\nu}^{2}M_{N} }{ 8\pi }\left( \frac{ M_{N} }{ T } \right),
	\end{equation}
requiring $n_{N}\Gamma \lesssim n_{l}H$ where $n_{N}$ ($n_{l}$) is the $N_{R}$ (lepton) number density leads to a washout avoidance condition:
	\begin{equation}
	\lambda_{\nu} \lesssim 10^{-7} \left( \frac{ g }{ 100 } \right)^{1/4} \left( \frac{ M_{N} }{ 10 \; \text{TeV} } \right)^{1/2}.
	\end{equation}

\item The mixing between the $N_{R}$ and $S_{3L}$ states is controlled by $f_{R3}= \sqrt{2}M_{N}/v_{\chi}$. Scattering $\Phi + l_{L} \to N_{R}/S_{3L} \to \eta_{1}^{\ast}+\overline{S_{2L}}$ has an approximate cross section,
	\begin{equation}
	\sigma \sim \frac{ (\lambda_{\nu}f_{23})^{2} }{ 8\pi } \frac{ M_{N}^{2} }{ (T^{2}+M_{N}^{2})^{2} }.
	\end{equation}
The scattering rate $\Gamma \sim \sigma T^{3}$ is inefficient compared to $H$ provided:
	\begin{equation}
	f_{R3} \lesssim 10^{-6} \left( \frac{ 1 }{ \lambda_{\nu}f_{23} } \right)\left( \frac{ g }{ 100 } \right)^{1/4}\left( \frac{ \Lambda_{\rm ISS} }{ 10 \; \text{TeV} } \right)^{1/2},
	\end{equation}
and for such values washout is avoided.

\item The Yukawa couplings $f_{23}$ and $f_{12}$ induce two body decays of the heaviest particle involved in the relevant interaction (such decays are kinematically allowed given $S_{2L}$, which has a radiative and hence negligible mass compared to the other particles, can appear in the final state). We estimate the decay rate as,
	\begin{equation}
	\Gamma \approx \frac{ (f_{i})^{2}M_{j} }{ 16\pi }\left( \frac{ M_{j} }{ T } \right),
	\end{equation}
where $f_{i}$ denotes the relevant Yukawa coupling and $M_{j}$ the mass of the heaviest particle. This leads to a washout avoidance condition of:
	\begin{equation}
	f_{i} \lesssim 10^{-6} \left( \frac{ g }{ 100 } \right)^{1/4}\left( \frac{ \Lambda_{\rm ISS} }{ 10 \; \text{TeV} } \right)^{1/2}.
	\end{equation}

\item Scattering processes $S_{2L} + \eta_{1}^{\ast} \to S_{1}/S_{1}^{c} \to \overline{S_{2L}}+\eta_{1}$ can be suppressed with small $f_{11}= M_{1}/\sqrt{2}v_{\chi}$. We estimate the cross section to be:
	\begin{equation}
	\sigma \sim \frac{ (f_{12})^{4} }{ 8\pi } \frac{ M_{1}^{2} }{ (T^{2}+M_{1}^{2})^{2} }.
	\end{equation}
This gives a washout avoidance condition:
	\begin{equation}
	f_{11} \lesssim 10^{-7} \left( \frac{ 1 }{ f_{12}  }\right)^{2}\left( \frac{ g }{ 100 } \right)^{1/2} \left( \frac{ \Lambda_{\rm ISS} }{ 10 \; \text{TeV} } \right)^{1/2}.
	\end{equation}

\item Washout may proceed through decays and inverse decays $\eta_{1}$ provided they are kinematically allowed. However, for a small enough $A$, and hence small enough mass splitting between $Re(\eta_{1})$ and $Im(\eta_{1})$, destructive interference leads to washout suppression. We estimate the decay rate as,
	\begin{equation}
	\Gamma \approx \frac{(|f_{23}|^{2}+|f_{12}|^{2})M_{I}}{16\pi}\frac{ M_{I} }{ T },
	\end{equation}
and the effective washout parameter to be,
	\begin{equation}
	 K_{\rm eff}=\left( \frac{ \Gamma }{ H }\Big|_{T={M_{I}}} \right) \cdot \delta^{2},
	\end{equation}
where $\delta = (m_{R}-m_{I})/\Gamma$. One finds this translates into a washout avoidance condition:
	\begin{equation}
	A \lesssim 10^{-4} \; \text{GeV} \; \sqrt{|f_{23}|^{2}+|f_{12}|^{2}} \left( \frac{ g }{ 100 } \right)^{1/4}\left( \frac{ \Lambda_{\rm ISS} }{ 10 \; \text{TeV} } \right)^{3/2}.
	\end{equation}
If such decays are not kinematically allowed washout can instead proceed through a scattering process mediated by off-shell $\eta_{1}$. We estimate the cross section as:
	\begin{align}	
	\sigma \sim & \frac{ (|f_{23}|+|f_{12}|)^{4} }{ 128\pi }\left(\frac{1}{ T^{2}+ m_{R}^{2}} -\frac{1} { T^{2}+  m_{I}^{2} }\right)^{2}T^{2} \\
		 = & \frac{ (|f_{23}|+|f_{12}|)^{4} }{ 128\pi }\left( \frac{ 2\sqrt{2}Av_{\chi} }{(T^{2}+ m_{R}^{2})(T^{2}+  m_{I}^{2}) } \right)^{2}T^{2}.
	\end{align}
This translates into a washout avoidance condition:
		\begin{equation}
		A \lesssim 10^{-2} \; \text{GeV} \left( \frac{ 1 }{ |f_{23}|+|f_{12}| } \right)^{2} \left( \frac{ g }{ 100 } \right)^{1/4}\left( \frac { \Lambda_{\rm ISS} }{ 10 \; \text{TeV} } \right)^{3/2}.
	\end{equation}

\end{enumerate}
Satisfying any of the above washout bounds means the BAU is safe. However, similarly to the Law/McDonald model, such small choices for the couplings ruins the ISS mechanism. 

\item $T_{\rm BG} < \Lambda_{\rm ISS}$

So as to ensure the preservation of the BAU, while allowing for the ISS mechanism, we may suppress washout by taking one or more of the exotic particle masses to be higher than $T_{\rm BG}$. This ensures that at $T_{\rm BG}$ the number density of that particle will be sufficiently Boltzmann suppressed for it to not play a role in washout. For decay processes the requirement on the mass scale is:
	\begin{equation}
	\frac{ \Lambda_{\rm ISS} }{ T_{\rm BG} } \gtrsim 42,
	\label{eq:oneboltzmannma}
	\end{equation}
while for scattering processes the constraint is relaxed to roughly,
	\begin{equation}
	\frac{ \Lambda_{\rm ISS} }{ T_{\rm BG} } \gtrsim 25.
	\label{eq:twoboltzmannma}	
	\end{equation}
This is again similar to the washout in the Law/McDonald model. For $T_{\rm BG} < T_{EW}$ the bounds of eqs. (\ref{eq:oneboltzmannma}) and (\ref{eq:twoboltzmannma}) do not apply as the sphalerons are no longer active.
\end{itemize}

\subsection{Resonant leptogenesis}
\label{sec:resonantleptogen}
For typical parameter choices, the lightest pseudo-Dirac pair of neutrinos, $N_{\alpha}$ will generate the BAU in resonant leptogenesis. There is an interplay between the $CP$ violation and washout, which both depend on $\mu$, the mass splitting of the $N_{\alpha}$ pair. So as to show that this model can generate the right order of magnitude for the parameters, let us take as a guide the example of \cite{Blanchet:2009kk}, which found the observed BAU can be generated with $\delta = \mu/\Gamma_{\alpha} \approx 10^{-5}$ ($\Gamma_{\alpha}$ is the width of the $N_{\alpha}$ and $\mu$ the mass splitting of the pair), heavy neutrino mass $M_{N \alpha}\sim\mathcal{O}(1)$ TeV, and Yukawa coupling $\lambda_{\nu}\approx 5\times10^{-2}$.

The scalar mass $m_{I}$ can be taken to a sufficiently high scale so that the other interactions of the ISS do not interfere with the generation of the BAU. Let us take $m_{I}\sim 50$ TeV and $v_{\chi}\sim A\sim M_{1} \sim 10$ TeV, then $M_{N} \sim\mathcal{O}(1)$ TeV can be achieved with $f_{R3}\sim 0.1$. Note that solutions of the Boltzmann equations taking into account decay and scattering processes show the BAU is generated at $T_{\rm BG} \approx M_{N \alpha}/10$, which allows the out-of-equilibrium condition for $N_{\alpha}$ to be met even with a relatively low $Z'$ mass: $M_{Z'} \gtrsim 2M_{N \alpha}$ \cite{Blanchet:2010kw}.

Choices of $\delta= 10^{-5}$,  $\lambda_{\nu}=0.05$, and $M_{N \alpha}=2$ TeV imply: $\mu = 0.2 \times 10^{-5}$ GeV. This is compatible with a radiative origin in the Ma model as, for these parameters, eq. (\ref{eq:massthreeapprox}) gives $\mu \approx (f_{12})^{2}(f_{23})^{2}\times5\times10^{-3}$ GeV. The light neutrino mass is $m_{\nu} \approx 10^{-2}$ eV.

In the $\delta \ll 1$ regime, the $CP$ asymmetry of $N_{\alpha}$ decays is \cite{Blanchet:2009kk},
	\begin{equation}
	\epsilon \simeq -\frac{ \lambda_{\nu}^{2} }{ 16\pi }\frac{ \mu \tilde{\mu} }{ (\mu^{2}+\tilde{\mu}^{2}+2\mu\tilde{\mu}\cos{\alpha}) } \sin{\alpha},
	\end{equation}
where $\alpha$ is a physical phase of the ISS mass matrix and $\tilde{\mu}$ is the Majorana mass $\tilde{\mu}\overline{N_{R}}(N_{R})^{c}$. In this model one typically expects $\tilde{\mu} \sim 10^{-3}\mu$, due to suppression by additional Yukawa couplings in the radiative generation of $\tilde{\mu}$, which allows for $\epsilon \sim 10^{-8}$. The washout factor is $K_{\rm eff} \approx 6$. Taking into account a sphaleron reprocessing and dilution factor, the final baryon number density to photon number density is, $\eta_{B}\sim 10^{-2}|\epsilon|$, and can easily be made to match the observed value $\eta_{B} \approx 6\times 10^{-10}$ \cite{Ade:2013lta} in this scenario.

A more detailed study finding the full form of the $3\times3$ matrices $m_{D}$ and $\mu$, and mass spectrum for the pseudo-Dirac pairs compatible with the observed oscillation parameters, BAU, and $SO(10)$ origin of the model is beyond the scope of this paper. 

\subsection{The $\rho$ parameter}
In the Ma model, the Higgs doublet responsible for EW symmetry breaking also carries $U(1)_{\chi}$ charge. This leads to a tree level correction to the $Z$ boson mass, and consequently to the $\rho$ parameter,
	\begin{equation}
	\rho \equiv \frac{ M_{W}^{2} }{ M_{Z}^{2}\mathrm{cos}^{2}\theta_{W} },
	\end{equation}
where the Weinberg angle is defined in terms of the weak isospin, $g$, and weak hypercharge, $g_{Y}$, coupling constants as $\tan{\theta_{W}}\equiv g_{Y}/g$. Experimental observations give a value $\rho-1=4^{+3}_{-4}\times10^{-4}$ \cite{PhysRevD.86.010001}. The masses of the neutral gauge bosons are given by,
	\begin{align}
	m_{\pm}^{2}=&\frac{ v_{W}^{2} }{ 8 } \Bigg\{ g^2+g_{Y}^{2}+16g_{\chi}^{2}\left[1+\frac{ v_{\chi}^{2} }{ v_{W}^{2} }\right] \nonumber \\
	&  \pm \sqrt{ \left(g^2+g_{Y}^{2}-16g_{\chi}^{2}\left[1+\frac{ v_{\chi}^{2} }{ v_{W}^{2} }\right]\right)^{2}+4[4g_{\chi}g_{Y}\sin{\theta_{W}}+4g_{\chi}g\cos{\theta_{W}}]^{2}} \Bigg\},
	\end{align}
so to remain within the $2\sigma$ limit of the $\rho$ parameter, one requires $v_{\chi}\gtrsim7.7$ TeV (given $g_{\chi} \sim g_{Y}$).\footnote{The constraint can also be satisfied with $g_{\chi} \lesssim g_{Y}/95$ and $v_{\chi}\lesssim5$ TeV. Given the GUT origin of the model, we consider such small choices for $g_{\chi}$ unattractive. Furthermore the interactions of the resultant light $Z'$ are constrained to be so weak by observations of Caesium transitions \cite{Wood:1997zq,Ginges200463,Bouchiat:2004sp}, that they would also lead to overclosure of the universe.}
The $Z-Z'$ mixing angle is given by,
	\begin{equation}
	\tan{2\phi_{Z}} = \frac{ 2(g_{\chi}g_{Y}\sin{\theta_{W}}+gg_{\chi}\cos{\theta_{W}}) }{ g^{2}+g_{Y}^{2}-4g_{\chi}^{2}(1+v_{\chi}^{2}/v_{w}^{2}) },
	\end{equation}
so one finds $|\phi_{Z}| \simeq (v_{w}/v_{\chi})^{2} \lesssim 10^{-3}$ given $v_{\chi} \gtrsim 7.7$ TeV. We will see the consequences of this for our DM candidate below.

So far we have dealt with mixing coming from the mass matrix. As the theory contains particles charged under both $U(1)_{Y}$ and $U(1)_{\chi}$, kinetic mixing between the respective field strength tensors will be generated radiatively. The full mass matrix for the $Z$ and $Z'$ bosons, taking into account both kinetic and mass mixing, can then be found \cite{Foot1991509}. By following this procedure we have checked that including such a term makes no qualitative difference to our discussion above. Furthermore one typically expects radiatively generated kinetic mixing to be small \cite{Baumgart:2009tn} and therefore negligible for mixing in this model.

There is also a competing limit from the LHC. For example, using the ATLAS limit on dilepton resonances \cite{ATLAS-CONF-2013-017} and a choice of $g_{\chi}=0.1$, one finds a lower bound of $M_{Z'}\gtrsim2.7$ TeV. Given that $M_{Z'}\simeq2g_{\chi}v_{\chi}$, this corresponds to $|\phi_{Z}|\lesssim6\times10^{-4}$. Larger $g_{\chi}$ means stronger collider limits on $M_{Z'}$. However, ATLAS and CMS currently only provide limits for $Z'$ resonances up to $3.5$ TeV \cite{ATLAS-CONF-2013-017,CMS-PAS-EXO-12-061}, and as the limit on the possible signal strength is rising for high mass regions, we do not pursue the details here. Suffice to say the bound on the mixing angle is approximately $|\phi_{Z}|\lesssim 10^{-3}$.

\subsection{Dark matter}
\subsubsection{DM candidate mass}
\begin{figure}[h]
\begin{center}
\includegraphics[width=450pt]{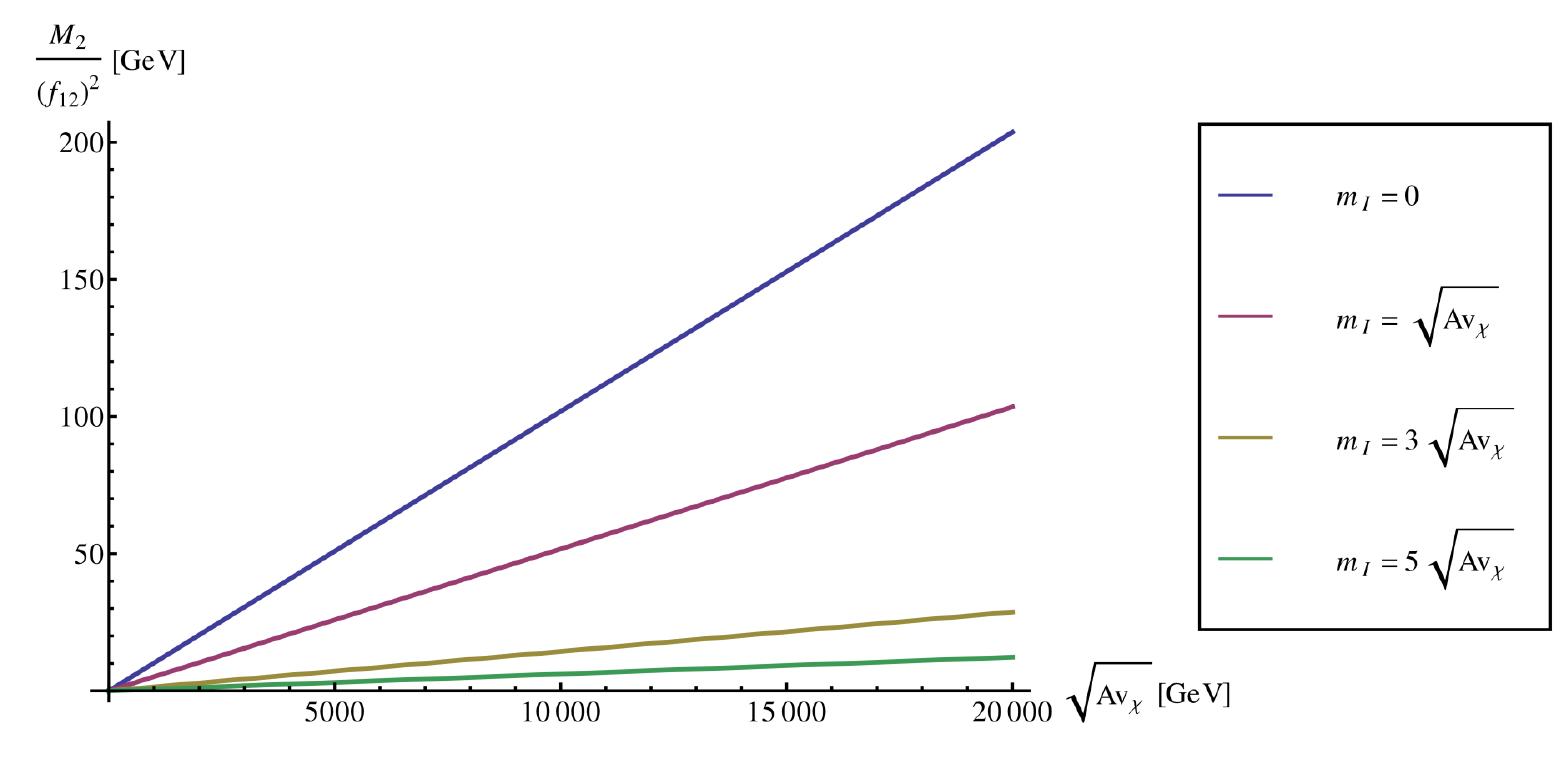}
\end{center}
\caption{$M_{2}/(f_{12})^{2}$ as a function of $\sqrt{Av_{\chi}}$, with $M_{1}=\sqrt{Av_{\chi}}$, for different choices of $m_{I}$.}
\label{fig:detuning}
\end{figure}

In the Ma radiative ISS, there exists an accidental $Z_{2}$ symmetry under which only $\eta_{1}$ and $S_{2L}$ transform. The lightest of these species is stable and forms a DM candidate. By examining the scalar potential in eq. (\ref{eq:morescalars}) we see that the masses of the real and imaginary components of $\eta_{1}$ after symmetry breaking are given by,
	\begin{align}
	m_{R}^{2} = (\mu_{\eta1}^{2}+\lambda_{1 \phi}v_{w}^{2}/2+\lambda_{12}v_{\chi}^{2}/2)^{2}+\sqrt{2}Av_{\chi} \\
	m_{I}^{2} = (\mu_{\eta1}^{2}+\lambda_{1 \phi}v_{w}^{2}/2+\lambda_{12}v_{\chi}^{2}/2)^{2}-\sqrt{2}Av_{\chi},
	\end{align}
where $A>0$ and for reasons of naturalness one typically expects $\mu_{\eta 1} \sim v_{\chi} \sim A$. The fermion $S_{2L}$ gains a radiative Majorana mass, $M_{2}$, at 1-loop order, and one typically expects,
	\begin{equation}
	m_{I} \gg M_{2},
	\end{equation}
if fine-tuning is absent. To provide a sense of the typical values for $M_{2}$, we have have plotted $M_{2}$ as a function of $\sqrt{Av_{\chi}}$, with $M_{1}=\sqrt{Av_{\chi}}$ for different choices of $m_{I}$ in figure \ref{fig:detuning}. One can see that as one moves away from the fine tuned limit $m_{I}=0$, the radiative mass $M_{2}$ decreases to be in the range,
	\begin{equation}
	M_{2} \approx (10^{-2}-10^{-3})(f_{12})^{2}v_{\chi}.
	\end{equation}

\subsubsection{Cold DM scenario}
\begin{figure}[h]
\begin{center}
\includegraphics[width=250pt]{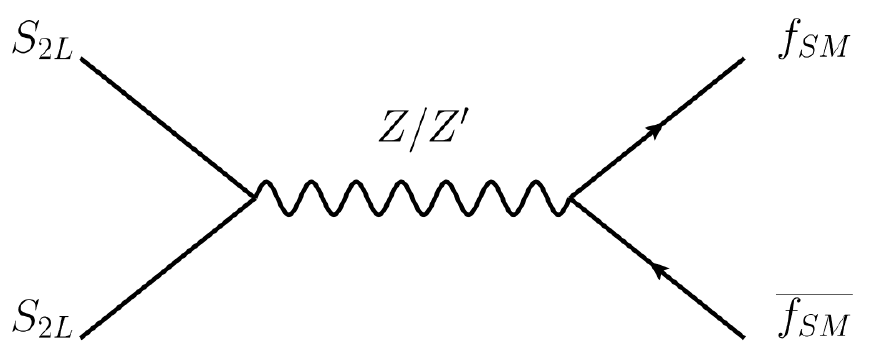}
\end{center}
\caption{Annihilation process of $S_{2L}$ which determines its relic density.}
\label{fig:colddarkmatter}
\end{figure}

Given the discussion of the $\rho$ parameter one requires $v_{\chi} \gg v_{w}$. We will discuss the DM relic abundance in light of this constraint here. Let us assume the $S_{2L}$ species have a thermal abundance in the early universe, as will be the case unless $g_{\chi}\lesssim10^{-8}\ll g_{Y}$. There is furthermore an absence of any interactions of the form $S_{2L}S_{2L}(\sigma)$ where $\sigma$ represents any of the scalar fields in the theory. As $S_{2L}$ is charged only under $U(1)_{\chi}$, it can annihilate only through the $Z'$, and due to mixing of the gauge bosons, through the $Z$ boson \cite{PhysRevD.41.1067} (see figure \ref{fig:colddarkmatter}). 

So as not to overclose the universe one requires the annihilation cross section to be larger than $\langle\sigma|v|\rangle_{DM}\approx1.9\times10^{-9}\;\mathrm{GeV}^{-2}$ \cite{Steigman:2012nb}. However annihilation through the $Z$ boson is suppressed by the small $Z-Z'$ mixing angle. For example, consider the $2M_{2} \lesssim M_{Z}$ regime. It suffices for our purposes to use a simple estimate for the annihilation cross section:
	\begin{equation}
	\langle\sigma|v|\rangle \approx \frac{ \sin^{2}{\phi_{Z}}G_{F}^{2}M_{2}^{2} }{ 2\pi },
	\end{equation}
where $\phi_{Z}$ is the $Z-Z'$ mixing angle and $G_{F}$ is the Fermi constant. Requiring $\langle\sigma|v|\rangle \gtrsim \langle\sigma|v|\rangle_{DM}$, and substituting for the mixing angle $\phi_{Z} \simeq (v_{w}/v_{\chi})^{2}$, one finds:
	\begin{equation}
	\label{eq:zone}
	M_{2} \gtrsim  \frac{ \sqrt{2\pi \langle\sigma|v|\rangle_{DM}} }{ G_{F} } \left( \frac{ v_{\chi} }{ v_{w} } \right)^{2} .
	\end{equation}
Given the constraint from the $\rho$ parameter, $v_{\chi}\gtrsim 7$ TeV, this demands $M_{2} \gtrsim 7.5$ TeV, which contradicts our original assumption of $2M_{2} \lesssim M_{Z}$. One also finds a too small cross section in the $2M_{2} \gtrsim M_{Z}$ regime. 

Annihilations occuring through the $Z'$ are suppressed by the large mass $M_{Z'}$. So annihilations through the $Z'$ are also too weak. 

Finally we note that the cross section will be much higher at a resonance point, where $2M_{2} \simeq M_{Z}$ or $2 M_{2} \simeq M_{Z'}$ \cite{PhysRevD.41.1067}, and this could avoid overclosure. However, this corresponds to a fine tuning of the DM mass.

From this discussion we see that for a simple cold DM scenario the $S_{2L}$ DM candidate overcloses the universe if we demand no fine tuning. One would have to go to more convoluted DM scenarios such as warm dark matter to perhaps find viable regions of parameter space. The simple cold DM case is ruled out.

\subsection{Concluding remarks on the Ma model}
The Ma radiative ISS model has several positive features for baryogenesis and washout avoidance. Resonant leptogenesis may be accommodated, as the mass splitting of the heavy neutrinos is generated before the EW phase transition, allowing the interplay between heavy neutrino decays and sphalerons crucial for leptogenesis. Also, washout due to the other $\Delta L$ interactions required for the radiative ISS can be suppressed with suitable choices of mass parameters. 

On the other hand we have also seen a drawback of the Ma model with respect to cold DM. We found that the constraints from measurements of the $\rho$ parameter translated into DM annihilations through the $Z$ and $Z'$ gauge bosons which are too weak to avoid overclosure. These simple cold DM scenarios are therefore ruled out for the Ma radiative ISS (apart for possibly small areas of parameter space corresponding to resonant enhancements). One would therefore have to go to more convoluted scenarios such as warm dark matter to find a suitable DM scenario.

\section{Conclusion}
Motivated originally by how new physics explanations for the baryon asymmetry, dark matter, and neutrino oscillations can overlap with each other, we began by looking at the properties of the inverse seesaw. We saw how the inverse seesaw mechanism offers various advantages, namely a scale low enough for direct experimental tests, and the possibility of resonant leptogenesis, or at least a suppressed washout rate if some other mechanism is responsible for the baryon asymmetry. 

The inverse seesaw mechanism does comes with a cost: the small lepton number violating parameter that must enter the mass matrix. Despite its technical naturalness, there may be a more appealing mechanism to generate it. One of the options  is a radiative origin. We therefore continued our investigation by focusing on two radiative inverse seesaw models.

It was found in the case of the Law/McDonald radiative inverse seesaw model, the DM abundance is easily explained with the Higgs portal mechanism, or the double Higgs portal. The advantages of the inverse seesaw with respect to baryogenesis, however, are lost due to the extra fields present which break lepton number explicitly with their interactions. In case of the Ma radiative inverse seesaw model, the situation is reversed: it nicely accommodates resonant leptogenesis, but does not feature a simple cold dark matter candidate.

\section*{Acknowledgements}
We thank S. S. C. Law and K. L. McDonald for their useful comments. IB was supported by the Commonwealth of Australia. NFB, KP, and RRV were supported by the Australian Research Council. KP was also supported by the Netherlands Foundation for Fundamental Research of Matter (FOM) and the Netherlands Organisation for Scientific Research (NWO). 

\bibliographystyle{JHEP}

\providecommand{\href}[2]{#2}\begingroup\raggedright\endgroup

\end{document}